# The most massive black holes on the Fundamental Plane of Black Hole Accretion

M. Mezcua,[1,2]⋆ J. Hlavacek-Larrondo,[1] J.R. Lucey,[3] M. T. Hogan,[3,4] A.C. Edge,[3] and B.R. McNamara[4]
[1]*Département de Physique, Université de Montréal, C.P. 6128, Succ. Centre-Ville, Montréal, QC H3C 3J7, Canada*
[2]*Institute of Space Sciences (IEEC-CSIC), Campus UAB, Carrer de Magrans s/n, E-08193 Barcelona, Spain*
[3]*Centre for Extragalactic Astronomy, Department of Physics, Durham University, South Road, Durham DH1 3LE, UK*
[4]*Department of Physics and Astronomy, University of Waterloo, Waterloo, ON, N2L 3G1, Canada*

**ABSTRACT**
We perform a detailed study of the location of brightest cluster galaxies (BCGs) on the fundamental plane of black hole (BH) accretion, which is an empirical correlation between a BH X-ray and radio luminosity and mass supported by theoretical models of accretion. The sample comprises 72 BCGs out to $z \sim 0.3$ and with reliable nuclear X-ray and radio luminosities. These are found to correlate as $L_X \propto L_R^{0.75\pm0.08}$, favoring an advection-dominated accretion flow as the origin of the X-ray emission. BCGs are found to be on average offset from the fundamental plane such that their BH masses seem to be underestimated by the $M_{BH} - M_K$ relation a factor ∼10. The offset is not explained by jet synchrotron cooling and is independent of emission process or amount of cluster gas cooling. Those core-dominated BCGs are found to be more significantly offset than those with weak core radio emission. For BCGs to on average follow the fundamental plane, a large fraction (∼ 40%) should have BH masses $> 10^{10}$ $M_\odot$ and thus host ultramassive BHs. The local BH-galaxy scaling relations would not hold for these extreme objects. The possible explanations for their formation, either via a two-phase process (the BH formed first, the galaxy grows later) or as descendants of high-z seed BHs, challenge the current paradigm of a synchronized galaxy-BH growth.

## 1 INTRODUCTION

The discovery over the past 20 years of local empirical scaling relations between the mass of supermassive black holes (SMBHs) and some of their host galaxy properties (stellar velocity dispersion, $\sigma$, bulge luminosity, $L_V$, and bulge mass, $M_{bulge}$; Magorrian et al. 1998; Ferrarese & Merritt 2000; Gebhardt et al. 2000; Marconi & Hunt 2003; Häring & Rix 2004; Gültekin et al. 2009a; Graham et al. 2011; see McConnell & Ma 2013, Kormendy & Ho 2013 and Graham 2016 for a review) constitutes a major breakthrough in galaxy evolution. The BH-galaxy scaling relations do not only provide a means to estimate the BH mass when direct measurements are not available (e.g., Volonteri & Reines 2016), but suggest the existence of a common SMBH-galaxy evolution in which star formation and stellar growth could be regulated by active galactic nucleus (AGN) feedback (e.g., Silk & Rees 1998; Di Matteo et al. 2005; Springel et al. 2005; Hopkins et al. 2005), hierarchical merging (e.g., Peng 2007; Jahnke & Macciò 2011) or secular evolution (e.g., Kormendy 2013).

The tightest BH-galaxy relations are found for galaxies with 'classical' bulges (formed by mergers; Kormendy & Ho 2013) and 'core-Sérsic' spheroids (with $M_{BH} \gtrsim 10^8$ $M_\odot$ and dominated by dry mergers; Graham 2012; Graham & Scott 2013), while barred galaxies, galaxies with 'pseudo'-bulges (dominated by secular evolution; Kormendy & Ho 2013), and 'Sérsic' spheroids (with $M_{BH} \lesssim 10^8$ $M_\odot$ and dominated by gas-rich processes; Graham 2012, Graham & Scott 2013) seem to follow a steeper relation with a larger scatter (e.g., Graham & Scott 2013). Yet, several types of sources have been found to be outliers of the scaling relations, both at the low- (e.g., Greene et al. 2008; Jiang et al. 2011; Graham & Scott 2013, 2015; Baldassare et al. 2015, 2017; see Mezcua 2017 for a review) and high-mass end (e.g., van den Bosch et al. 2012; Bogdán et al. 2012; Emsellem 2013; Yıldırım et al. 2015; Ferré-Mateu et al. 2015, 2017; Scharwächter et al. 2016; Walsh et al. 2015, 2016; Secrest et al. 2017). This is the case for some of the most luminous and massive galaxies in the local Universe – brightest cluster galaxies (BCGs) – for which dynamical BH mass measurements are available (the BCGs in A3565 and A1836, Dalla Bontà et al. 2009; NGC 6086 in A2162, McConnell et al. 2011b; NGC 3842 in A1367 and NGC 4889 in Coma, McConnell et al. 2011a; and M87[1] in Virgo, Gebhardt et al. 2011). These sources tend to be overmassive with respect to the $M_{\rm BH}-\sigma$ relation, but appear more in agreement, with a large scatter, with the $M_{\rm BH}-L_{\rm V}$ and $M_{\rm BH}-M_{\rm bulge}$ relations (e.g., Lauer et al. 2007; McConnell et al. 2011a; McConnell & Ma 2013).

⋆ E-mail: marmezcua.astro@gmail.com

BCGs are giant elliptical galaxies that reside near the gravitational center of galaxy clusters, at the intersection of cosmological dark matter filaments. They are located in very dense environments and thus subject to conditions very different to those of field/isolated galaxies. The intracluster medium in the core of some galaxy clusters known as 'cool-core clusters' (CCs) have very short central cooling times ($t_{\rm cool}$ < 5 Gyrs) and present highly peaked X-ray surface brightness profiles. CCs are thus expected to have high cooling rates at their centers (i.e., a cooling flow rate of $\sim 10^2-10^3$ $M_\odot$ yr$^{-1}$); yet, only a few percent of this cooling is observed in the form of cold gas and star formation at $z\sim0$ (e.g., O'Dea et al. 2008; the so-called 'cooling flow' problem; Fabian 1994). AGN feedback is thought to provide the necessary heating to counteract the cooling through radiatively inefficient accretion in a 'radio-mode' (e.g., see reviews by McNamara & Nulsen 2007, 2012; Fabian 2012; Heckman & Best 2014). The mechanical heating imparted by the AGN jets in the radio-mode feedback is favored by the observation of large jet-inflated cavities in the X-ray emitting intracluster medium (e.g., Fabian et al. 2000; McNamara et al. 2000; Hlavacek-Larrondo et al. 2012a). At $z\geq0.6$, McDonald et al. (2016) find that most of the BCGs in their sample have star formation rates exceeding 10 $M_\odot$ yr$^{-1}$ (see also Webb et al. 2015a,b). These 'star-forming' BCGs are hosted by unrelaxed, non-cool core clusters (NCCs) lacking a central enhancement in their X-ray profiles, which suggests that galaxy mergers instead of cooling flows might be triggering star formation at high $z$. A strong redshift evolution was hinted for a sample of 32 BCGs located in clusters with clear X-ray cavities (i.e., mostly CCs with strong radio-mode feedback) and whose nuclear X-ray luminosity has fainted by a factor $\sim$ 10 since $z\sim0.6$ (Hlavacek-Larrondo et al. 2013). This suggests that the fraction of radiatively efficient BCGs might be higher at $z\sim0.6$ than in the local Universe and that BCGs with radio-mode feedback might have transitioned from the analogous low/hard state of X-ray binaries to a quiescent state over the last 5 Gyrs (Hlavacek-Larrondo et al. 2013).

How BCGs form and evolve is thus not yet clear. They might have gone through major mergers in the past (e.g., De Lucia & Blaizot 2007; Webb et al. 2015a,b; McDonald et al. 2016) but are currently subject to powerful AGN feedback that counteracts most of the expected cooling (e.g., McNamara et al. 2000; Hlavacek-Larrondo et al. 2012a). Do BCGs then follow the same co-evolutionary galaxy/BH growth as non-BCG AGN? Which is the behavior of the BH-galaxy scaling relations in the high-mass end, where BCGs are located? One powerful means to probe this is to investigate the location of BCGs on the fundamental plane of BH accretion, which is a correlation between nuclear X-ray luminosity, core radio luminosity and BH mass that extends from stellar-mass to SMBHs and that thus unifies BHs across all mass scales (e.g., Merloni et al. 2003; Falcke et al. 2004; Körding et al. 2006; Gültekin et al. 2009b; Plotkin et al. 2012; Xie & Yuan 2017).

[1] M87 is located at the center of the Virgo cluster but it is not the BCG; the brightest galaxy is M49.

The only dedicated study to the location of BCGs on the fundamental plane was performed by Hlavacek-Larrondo et al. (2012b), who studied a sample of 18 BCGs residing in strong cooling flow clusters (with $t_{\rm cool}$ < 3 Gyrs and X-ray luminosities $\geq 10^{45}$ erg s$^{-1}$; Hlavacek-Larrondo & Fabian 2011) with no detectable X-ray nucleus. In these strong CCs, the AGN should provide a large amount of mechanical feedback ($L_{\rm mechanical}$ > $10^{45}$ erg s$^{-1}$) to counteract the cooling of the surrounding gas. These AGN are expected to accrete substantially and thus to be detected as an X-ray point source. Yet, all the strong CCs with a central radio source indicating the presence of an AGN studied by Hlavacek-Larrondo & Fabian (2011) lack a detectable X-ray nucleus at their center. If the BCGs in these strong CCs had $M_{\rm BH}$ > $10^{10}$ $M_\odot$ (i.e., were 'ultramassive'), they would radiate inefficiently and thus should not necessarily show nuclear X-ray emission (Hlavacek-Larrondo & Fabian 2011). The existence of these ultramassive BHs is predicted from theoretical models (e.g., Inayoshi & Haiman 2016; King 2016) and a few of them have been found observationally based on dynamical BH mass measurements (e.g., McConnell et al. 2011b; the record-holder is the BCG NGC 4889, with $M_{\rm BH}=2.1^{+1.6}_{-1.6}\times10^{10}$ $M_\odot$, McConnell et al. 2011a). Their presence in strong CCs could thus explain the lack of an X-ray point source (although see Hlavacek-Larrondo & Fabian 2011 for alternative explanations such as a very high spin or obscuration).

Using low-resolution radio data, Hlavacek-Larrondo et al. (2012b) found that the sample of strong CCs from Hlavacek-Larrondo & Fabian (2011) is on average offset from the fundamental plane such that the BH masses of their BCGs are underestimated by the BH-galaxy correlations by a factor 10. This is, if the BCGs in strong CCs follow the fundamental plane, then they host ultramassive BHs with $M_{\rm BH}$ > $10^{10}$ $M_\odot$. Another possibility is that their X-ray luminosities are underestimated due to synchrotron cooling, or that their radio luminosities are overestimated due to the low resolution of the radio data (Hlavacek-Larrondo et al. 2012b). How does this extrapolate to CCs (with cluster X-ray luminosites < $10^{45}$ erg s$^{-1}$) and NCCs? CCs are less dynamically disturbed than NCCs and their BCGs have a higher gas supply; the BCGs in CCs are thus expected to host more massive BHs than those in NCCs. Is this reflected by CC-BCGs being more offset from the fundamental plane than NCC-BCGs, or do BCGs sit on the fundamental plane independently of the amount of cluster central cooling gas? Could it be that BCGs in general do not follow the fundamental plane (or that they follow a different one) because their BH is in quiescence instead of in a low/hard state (Hlavacek-Larrondo et al. 2013) or because their BH operate differently than those of AGN (i.e., because of being surrounded by a substantial amount of hot dense gas)?

To investigate all these possible scenarios and ultimately better understand the formation and growth of BCGs, we perform the first systematic study of BCGs using the fundamental plane of BH accretion. For this we use a large sample of 72 BCGs with high-resolution radio observations and *Chandra* data (both detections and upper limits on the nuclear X-ray emission) that includes CCs as well as NCCs. As reported and discussed in Sect. 4, we find that BCGs sit on average above the fundamental plane and that this offset is stronger for those BCGs with core-dominated radio emission. The description of the BCG sample and data analysis are detailed in Sects. 2 and 3, respectively. Final conclusions and implications of our findings are given in Sect. 5. Throughout the paper we adopt a $\Lambda$CDM cosmology with $H_0$ = 70 km s$^{-1}$ Mpc$^{-1}$, $\Omega_\Lambda$ = 0.73 and $\Omega_m$ = 0.27.

## THE FUNDAMENTAL PLANE OF BLACK HOLE ACCRETION

The fundamental plane is an empirical correlation between 2-10 keV nuclear X-ray luminosity, 5 GHz nuclear radio luminosity and BH mass found for stellar-mass BHs and SMBHs (e.g., Merloni et al. 2003; Falcke et al. 2004; Körding et al. 2006; Gültekin et al. 2009b; Plotkin et al. 2012; Xie & Yuan 2017). It extends over six orders of magnitudes in BH mass and has also been observed in the intermediate-mass BH regime (Gültekin et al. 2014). The correlation is supported by theoretical models of accretion that predict the existence of a disk-jet coupling mechanism: the radio luminosity is a proxy of the (optically-thick) synchrotron emission produced by the jet, while the X-ray emission is attributed to either coronal emission from the accretion flow (the 'disk/jet model'; e.g., Yuan et al. 2005; see Yuan & Narayan 2014 for a review) or to optically-thin synchrotron emission near the base of the jet (the 'synchrotron/jet model'; e.g., Falcke et al. 2004; Körding et al. 2006; Plotkin et al. 2012; Wang & Dai 2017). The correlation constitutes a strong (if not the strongest) unification of BHs, proving that they are governed by the same accretion physics independently of their mass. The fundamental plane correlation that includes the largest and most diverse sample of objects is that from Merloni et al. 2003: $\log L_{\rm R} = (0.60 \pm 0.11)\log L_{\rm X} + (0.78 \pm 0.10)\log M_{\rm BH} + (7.33 \pm 4.06)$ where $L_{\rm R}$ is the 5 GHz radio luminosity, $L_{\rm X}$ the 2-10 keV X-ray luminosity, and it has a scatter of 0.88 dex in $L_{\rm R}$ (0.62 dex perpendicular to the plane). Their sample includes eight X-ray binaries (XRBs) and ∼100 AGN, among which there are low-luminosity AGN (LLAGN), Seyfert galaxies, low-ionization nuclear emission regions (LINERs), radio galaxies, and radio-loud and radio-quiet quasars. They include both flat and steep radio sources, hence the origin of the radio emission is unclear (it could include both core and radio lobe emission). They exclude BL Lac objects. The BH masses come from direct (e.g., reverberation mapping) and indirect (the $M_{\rm BH} - \sigma$ relation) methods and their sample includes different accretion rates. By including only nuclear radio sources with dynamical mass measurements (i.e. excluding reverberation BH masses) and with 5 GHz peak radio emission (18 sources in total compared to the more than 100 of Merloni et al. 2003), Gültekin et al. (2009b) reduce the scatter to 0.7 dex.

If the X-ray and radio emission arise predominantly from the BH jet, then only sources in the low/hard X-ray state and thus dominated by jet emission (Fender 2001), or with sub-Eddington accretion ($< 10^{-2}$; equivalent to the low/hard state) should be included in the fundamental plane (e.g, Falcke et al. 2004; Körding et al. 2006). Körding et al. (2006) found that those sources accreting in the sub-Eddington regime follow the fundamental plane more tightly, while the inclusion of sources in a high state, with a strong contribution from the accretion disk, increases the scatter of the correlation. Considering a sample of only objects in the low/hard state, with flat/inverted radio spectra and 5 GHz radio luminosities (five XRBs and ∼50 AGN - which include LLAGN, radio galaxies, and BL Lacs, thought to host sub-Eddington BHs and to be jet-dominated; e.g., Falcke et al. 2004; Ghisellini et al. 2009), Körding et al. (2006) found a best-fitting relationship: $\log L_{\rm X} = (1.41 \pm 0.11)\log L_{\rm R} - (0.87 \pm 0.14)\log M_{\rm BH} - (5.01 \pm 3.20)$ with a intrinsic scatter of 0.38 dex, which they reduce to 0.12 dex when considering only XRBs and LLAGN. Gültekin et al. (2009b) also reduce the scatter from 0.7 dex (when considering different accretion rates) to 0.25 dex when considering only eight sources with low accretion rates (i.e., LINERs and LLAGN), supporting further that sources with high Eddington ratios may not be included in the fundamental plane.

The tightest correlation is that obtained by Plotkin et al. (2012), who use the same parent sample limited to sub-Eddington accretion rates as Körding et al. (2006) but apply a Bayesian technique that allows them to account for correlated measurement errors between radio and X-ray luminosities and to measure the intrinsic scatter directly from the data. They derive a relationship that is the most robust and accurate for BHs with sub-Eddington accretion and with flat/inverted radio spectra: $\log L_{\rm X} = (1.45 \pm 0.04)\log L_{\rm R} - (0.88 \pm 0.06)\log M_{\rm BH} - (6.07 \pm 1.10)$ with an intrinsic scatter about the plane of 0.07 dex. While Merloni et al. (2003) derive a general expression that covers all accretion rates and where X-rays are due not only to optically-thin synchrotron emission but also inverse Compton (corona-dominated) emission, the X-ray emission in the correlations of Körding et al. (2006) and Plotkin et al. (2012) is dominated by synchrotron radiation from the jet (the synchrotron/jet model). The Körding et al. (2006) and Plotkin et al. (2012) correlations should thus not be applied to BHs with high accretion rates and dominated by coronal X-ray emission.

Radiative cooling can also strongly affect the fundamental plane correlations if the X-ray emission originates predominantly from the jet (e.g., Falcke et al. 2004; Körding et al. 2006; Plotkin et al. 2012). For those SMBHs more massive than $10^8$ $M_\odot$, the frequency at which the synchrotron emitting electrons suffer from significant cooling is below the X-ray band; the X-ray flux of these sources is thus smaller than what would be expected from an uncooled jet. Synchrotron cooling is thus a concern for the most massive BHs, as those hosted by BCGs. To address it, Körding et al. (2006) and Plotkin et al. (2012) extrapolate the X-ray luminosities for the most massive sources in their sample (e.g., BL Lacs) from a lower-frequency band (i.e. the optical) to ensure that optically-thin and uncooled jet synchrotron emission is used across the whole mass scale. However, synchrotron cooling is not taken into account in other correlations (i.e. the Merloni one), as no a priori assumption on the dominant radiative mechanism is made and the origin of the X-ray emission of their sources is unclear (it is probably a mixture of jet-dominated, e.g., the LLAGN, and corona-dominated, e.g., the radio-loud quasars; Plotkin et al. 2012; with the coronal origin most likely dominating over the jet model; Merloni et al. 2003). Note that Merloni et al. (2003) do not include BL Lacs in their sample, hence very few of their sources are expected to emit synchrotron-cooled X-ray radiation.

In this paper we will show the location of BCGs on (1) the correlation of Merloni et al. (2003), as this has the broadest types of objects, the broadest scatter, is not strongly affected by synchrotron cooling, and is independent of accretion rate and origin of the X-ray and radio emission, and (2) on that from Plotkin et al. (2012), as this is, together with that of Körding et al. (2006), the one that should better match our sample of BCGs (with sub-Eddington accretion and flat/inverted radio spectra), includes synchrotron cooling, and has the least scatter. The behavior of BCGs on other fundamental planes (e.g., Körding et al. 2006; Gültekin et al. 2009b; Xie & Yuan 2017) will be also discussed.

## 2 SAMPLE SELECTION

The sample of BCGs included in this paper is drawn from Hogan et al. (2015a), who performed a multifrequency radio study of BCGs drawn from an X-ray selected parent sample of 720 clusters. The 5 GHz core radio flux could be directly measured for those sources with available very long baseline interferometry (VLBI) radio observations, while for the remaining sources Hogan et al. (2015a) used the radio spectral energy distribution (SED) to decompose the BCG radio emission into a core (flat radio emission attributed to active, current AGN activity) and non-core (steep radio emission that traces more diffuse and aged lobe emission) component. Morphology, extension at other frequencies, and variability (indicative of the presence of a currently active core) were taken into consideration so that core radio fluxes could be reliably determined. For those sources for which the core component dominated the observed flux, that component was taken a measurement of the core component and a limit was placed on the non-core component by extrapolating with a steep spectral index ($\alpha_{\rm non-core}$ = 1.0) from the lowest observed frequency (see Hogan et al. 2015a for further details). Since there are some strong core-dominated sources in Hogan et al. (2015a) where a weak non-core component is still detected, we use the ratio $f_{\rm core}/f_{\rm non-core} \geq 0.1$, where $f_{\rm core}$ is the 5 GHz core radio flux and $f_{\rm non-core}$ the 1 GHz non-core radio flux (extrapolated with $\alpha_{\rm non-core}$ = 1.0), as an indication of the core dominance of source. We then define 'core-dominated' BCGs as having $f_{\rm core}/f_{\rm non-core} \geq 0.1$, and 'weak-core' BCGs (with a weak core plus extended radio emission) as having $f_{\rm core}/f_{\rm non-core} < 0.1$.

Hogan et al. (2015a) also classified the parent clusters into CC and NCC based on the presence of (predominantly $H_{\alpha}$+ [NII]) optical emission lines around the BCG. We adopt this same classification throughout this paper, distinguishing between CC and NCC clusters and between core-dominated and weak-core BCGs. We searched in the *Chandra* archive for available X-ray data for the 129 BCGs with core radio emission (i.e. core-dominated and weak-core sources) from Hogan et al. (2015a), finding that 85 sources had been observed by *Chandra* for more than 2 ks. We derive the 5 GHz core radio luminosity for these 85 sources. Because the 5 GHz radio fluxes come either from direct VLBI measurements or robust SED decomposition, we can be confident that the core radio luminosities are not overestimated.

## 3 DATA ANALYSIS

### 3.1 *Chandra* data reduction

The *Chandra* data for the 85 clusters with BCG radio emission were processed and analyzed using the CIAO software version 4.7 and CALDB4.6.9. For those sources with more than one *Chandra* observation, we selected either the most recent dataset or that with the largest exposure time. In some cases for which the exposure times where very short, we merged the level 2 event files of several observation IDs (e.g., A1795; see Table 2). The level 2 event files were filtered to the 0.5–7 keV energy band and background lightcurves were extracted of a field that excludes the chip were the BCG is located. Lightcurves were filtered using the lc_sigma_clip routine. The flux of the BCG in each cluster was then estimated photometrically and spectroscopically following the same procedures as in e.g., Hlavacek-Larrondo & Fabian (2011) and Russell et al. (2013).

*Photometric X-ray flux.* The location of the BCG nuclear X-ray emission was defined as a square region of 4 pixels × 4 pixels around the point of maximum brightness in the hard (3–7 keV) band image of the cluster (in the 0.5–7 keV band in the case of non-detection in the hard band). A square annulus region around that maximum location of 6 × 6 inner pixels$^2$ and 8 × 8 outer pixels$^2$ was taken as the background. We considered the nuclear X-ray emission as a detection if the counts in the square region of 4 pixels × 4 pixels are at least 3$\sigma$ above the background in the hard band. The net number of counts of the nuclear BCG X-ray emission was calculated by subtracting the background counts scaled to the same number of pixels as the source region. The noise in each region was estimated as $\sigma = \sqrt{N}$ (where $N$ is the count number) considering Poisson statistics, and error propagation was applied to estimate the 1$\sigma$ error on the net counts. The net counts were also corrected for the 90% fraction of the point spread function that falls in the 1 arcsec region. The net count rates were converted to flux and luminosity in the 2–10 keV band with PIMMS, using an unabsorbed power-law model with no intrinsic absorption, photon index $\Gamma$ = 1.9 and Galactic column density from COLDEN[2] (see Table 2).

*Spectroscopic X-ray flux.* The background cluster emission might be significantly undersubtracted by the photometric approach, specially in clusters with strong central surface brightness peaks. To obtain a better estimate of the nuclear X-ray luminosity than that provided by the count rate, we investigate the presence of a non-thermal component within the X-ray spectrum of the nuclear source. For this we use a circular region of 1 arcsec radius around the location of the nuclear X-ray source defined in the photometric method. To estimate the properties (temperature and abundance) of the thermal component of the cluster, we use as background region an annulus within an inner radius of 2 arcsec and outer radius 3 arcsec. Both regions were background-subtracted using a big region (of 250 arcsec) away from the cluster emission and loaded in *Sherpa* for consequent spectral fitting. We first keep the Galactic column density frozen and fit the annulus region with an absorbed (Galactic) thermal model (xsphabs ∗ xsapec in *Sherpa*) where the temperature, abundance, and normalization were free to vary. We used $\chi^2$ statistic with the Gehrels variance function (chi2gehrels, the default statistic in *Sherpa*). We extrapolate the temperature obtained in the annulus to the X-ray source region (circular region of 1 arcsec) using the relation $T = ar^b$, where $T$ is the temperature, $r$ is the radius and $b \sim 0.3$ (Voigt & Fabian 2004). We then use this extrapolated temperature, the abundance obtained in the thermal model (which is not expected to significantly vary from a radius of 2–3 arcsec to $r$ = 1 arcsec) and power-law index of 1.9 to fit the nuclear emission with an absorbed power-law + thermal model (xsphabs ∗ (powlaw1d + xsapec) model in *Sherpa*). The normalization of the thermal and non-thermal components were kept free with the only constraint that the normalization of the thermal component could not be less than that of the annulus region scaled for the same pixel number. Additional intrinsic absorption of the power-law was also allowed (xszphabs). The photon index was free to vary; however, in most cases we had to freeze the photon index and the column density to 1.9 and the Galactic values, respectively, in order to obtain a good fit (see Table 2). The unabsorbed 2-10 keV flux of the non-thermal component and 1$\sigma$ uncertainties were derived using the sample_flux *Sherpa* function. For eight of the target sources we failed to perform a proper photometric and spectroscopic fitting (e.g., because of the cluster being near the edge of chip or pile-up effects). We use the X-ray luminosities in the 2-10 keV band derived from the spectral fitting of the remaining 77 BCGs for further analysis.

[2] http://cxc.harvard.edu/toolkit/colden.jsp.

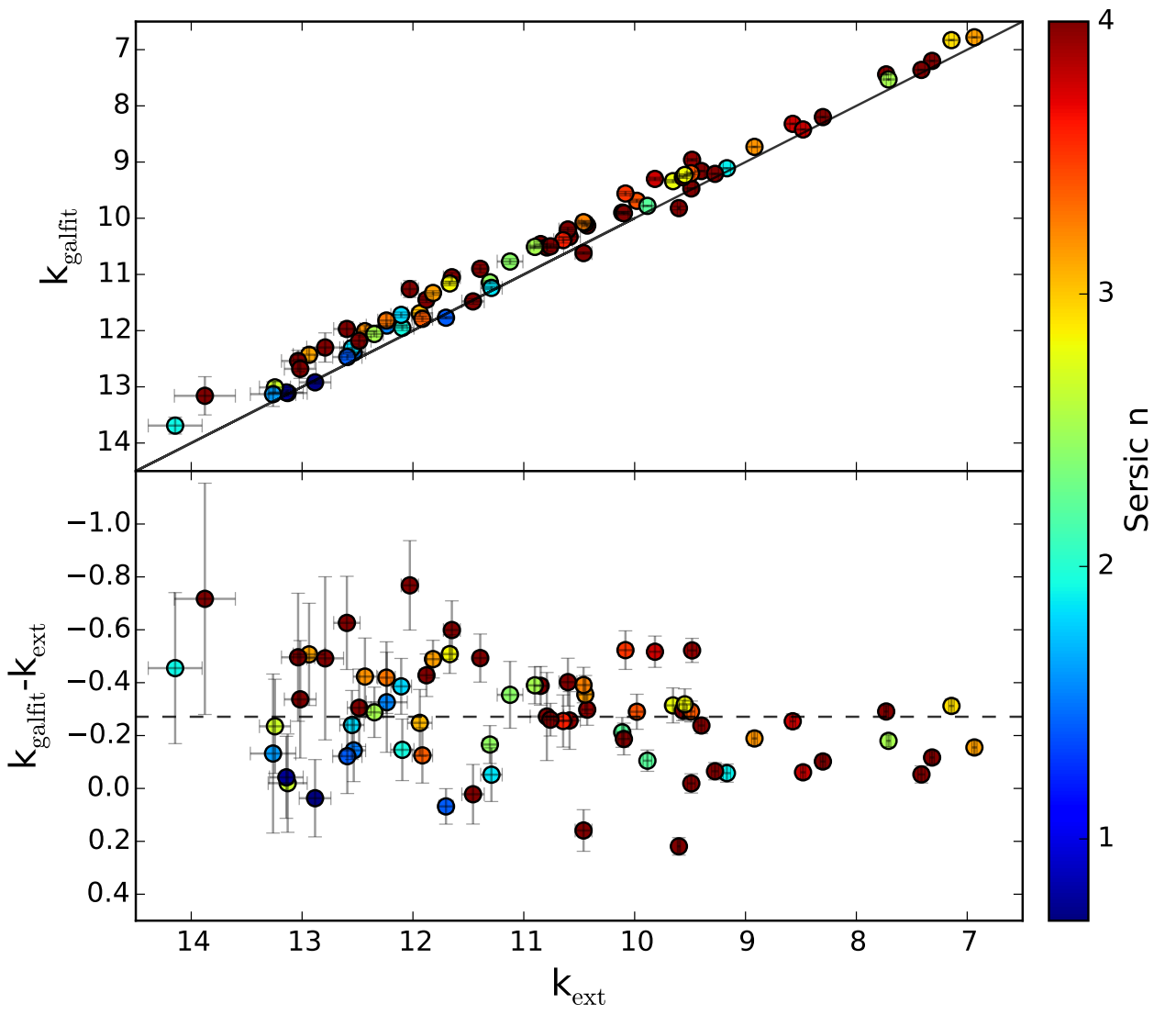

**Figure 1.** *Top*: Apparent $K$-band magnitudes derived from the GALFIT fit of a Sérsic profile with $n < 4$ versus total magnitudes from the 2MASS XSC. The one-to-one correlation is shown with a black line. *Bottom*: Difference between GALFIT magnitude and XSC magnitude versus XSC magnitude. The mean offset of -0.3 is shown with a dashed line.

### 3.2 K-band photometry and BH masses

To calculate the BH masses of the BCGs we use the $M_{\rm BH} - M_{\rm K}$ correlation from Graham & Scott (2013) found for a sample of 24 core-Sérsic galaxies:

$$\log M_{\rm BH} = (9.05 \pm 0.12) + (-0.44 \pm 0.08)(M_{\rm K} + 25.33) \qquad (1)$$

where $M_{\rm K}$ is the Two Micron All Sky Survey (2MASS) $Ks$-band spheroid magnitude and the scatter in log $M_{\rm BH}$ is of 0.44 dex. From now on we will refer to the 2MASS $Ks$-band filter as $K$-band. The $M_{\rm K}$ used by Graham & Scott (2013) are extracted from the 2MASS Extended Source Catalog (XSC; Jarrett et al. 2000). The total apparent magnitudes (or extrapolated magnitudes, *k_m_ext*) tabulated in the XSC are derived from a Sersic (1968) fit with $n < 1.5$ to the surface brightness profile (Jarrett et al. 2003; Lauer et al. 2007), which is known to underestimate the actual magnitudes by -0.33 mag (e.g., Schombert & Smith 2012; van den Bosch 2016). Graham & Scott (2013) are aware of this offset and correct for it in Eq. 1.

We derive $K$-band Sérsic magnitudes for the BCGs from the 2MASS image "postage stamps". A model two-dimensional Gaussian point-spread function (PSF) image was derived from stars on the parent 2MASS data tile. GALFIT (Peng et al. 2002, 2010) was used with the galaxy image and model PSF image as inputs to find the best-fitting two-dimensional Sérsic model. In the fitting we limit the Sérsic index to be less than 4, which we find is the best compromise for recovering the total extent of the surface brightness profile without having large residuals (e.g., see also van den Bosch 2016). For two sources (RXJ2250.0+1137 and RXJ2214.7+1350) the 2MASS photometry is not reliable because of the galaxy being too close to a tile corner in one case and being a close pair of galaxies in the other, while the sources MACS0159.8-0850, MACSJ0547.0-39, and RXJ1347.5-1144 are not bright enough for the 2MASS photometry to be performed. We thus remove these five sources from our sample. The final sample of BCGs that constitutes the focus of this paper contains thus 72 sources. Their redshifts span from $z = 0.006$ to $z = 0.29$ (see Table 2). We note that for A2355 its 2MASS counterpart has a redshift (2MASXJ21351874+0125269; $z = 0.2306$) that differs from that in the parent cluster sample of Hogan et al. (2015a). The most recent redshift reported for this source and used here is $z = 0.2310$ (Planck Collaboration et al. 2015).

We compare the $K$-band magnitudes derived from the fit of a $n < 4$ Sérsic profile with those $K$-band magnitudes directly extracted from the 2MASS $K$-band photometry in Fig. 1. As expected, the XSC magnitudes are significantly fainter than our fitted magnitudes. The offset between the two increases for the faintest sources, for which the Sérsic index is also less constrained and the uncertainties are large (Fig. 1, bottom panel). We find an average offset between the GALFIT and XSC $K$-band magnitudes of -0.3, in agreement with that found by previous authors (e.g., Schombert & Smith 2012; van den Bosch 2016). The $K$-band magnitudes derived from GALFIT are $K$-corrected using the total $J$-$K$ color from 2MASS and following the prescription of Chilingarian et al. (2010)[3], and corrected for Galactic extinction using the extinction law from Schlegel et al. (1998) and the coefficients from Cardelli et al. (1989).

We calculate the BH masses using the Graham & Scott (2013) correlation (Eq. 1) and the $K$-band magnitudes derived from GALFIT. The errors on the BH mass are obtained from propagation of the errors of each of the terms inEq. 1. We then plot the location of the BCGs on the fundamental planes of Merloni et al. (2003) (Fig. 2) and Plotkin et al. (2012) (Fig. 3) making a distinction in between CC/NCC clusters (left panels), and core-dominated/weak-core BCGs (right panels). In Fig. 2 we also include 80 individual SMBHs from Merloni et al. (2003) with direct (dynamical, e.g., stellar and gas kinematics) or indirect (from the $M_{\rm BH} - \sigma$ relation) BH mass measurements. These include the BCGs Cygnus A, NGC 1275 in Perseus, NGC 6166 in A2199 (also included in our sample, see Table 3, and whose BH mass in Merloni et al. 2003 of log $M_{\rm BH}$=9.19 $M_\odot$ is estimated from the stellar velocity dispersion), and the central galaxy M87 (or NGC 4486) in Virgo.

## 4 RESULTS AND DISCUSSION

Of the 72 BCGs included in our sample and plotted in Figs. 2-3, 31 do not have a detected X-ray nucleus in the hard band and their X-ray luminosities are thus taken as upper limits (shown with arrows in Figs. 2-3). Considering that these sources could lie more leftward, the BCGs appear to lie mostly on and above the correlations of Merloni et al. (2003) and Plotkin et al. (2012).

Hlavacek-Larrondo et al. (2012b) found that their sample of 18 BCGs with non X-ray detections sit on and above the fundamental plane of Merloni et al. (2003) so that their radio luminosities seem to be too high compared to their X-ray luminosities and BH masses. This is now also seen for our sample of 72 BCGs, which quadruplicates in size that of Hlavacek-Larrondo et al. (2012b) and includes 41 BCGs with a detected X-ray nucleus. The offset is even more striking when considering the correlation of Plotkin et al. (2012), as there *all* the BCGs except for one lie *above* the plane.

To probe the significance of these potential offsets we follow the same procedure as in Hlavacek-Larrondo et al. (2012b) and use a Monte Carlo technique to derive the average mass offset log $\Delta M_{\rm BH}$ that would be needed for the BCGs to follow the fundamental plane; this is, we determine which mass offset they need, on average, to lie above and below the plane. We assume that the measurements of $L_{\rm X}$, $L_{\rm R}$ and $M_{\rm BH,K}$ used in the fundamental plane correlations are independent and that they follow a Gaussian distribution based on their values and associated uncertainties. For each BCG we assign 100 random variables to the distribution of $L_{\rm X}$, 100 random variables to the distribution of $L_{\rm R}$, and 100 random variables to the distribution of $M_{\rm BH,K}$. We then calculate the log $\Delta M_{\rm BH}$ needed for the BCGs to follow the fundamental plane based on the distribution of $100^3$ possibilities over the 72 BCGs. The final mass offset and its error are taken as the median value and standard deviation, respectively, of the log $\Delta M_{\rm BH}$ distribution (see Table 1). The probability that the mass offset is larger than zero is also derived (in percentage and in $\sigma$; Table 1). The results do not change significantly if the mean is instead taken nor if the number of random variables is increased to e.g., 500. Since some of the X-ray luminosities are upper limits and not detections, the median values obtained in the Monte Carlo exercise should be taken as the minimum BH mass offset needed for the BCGs to sit on the fundamental plane.

[3] http://kcor.sai.msu.ru.

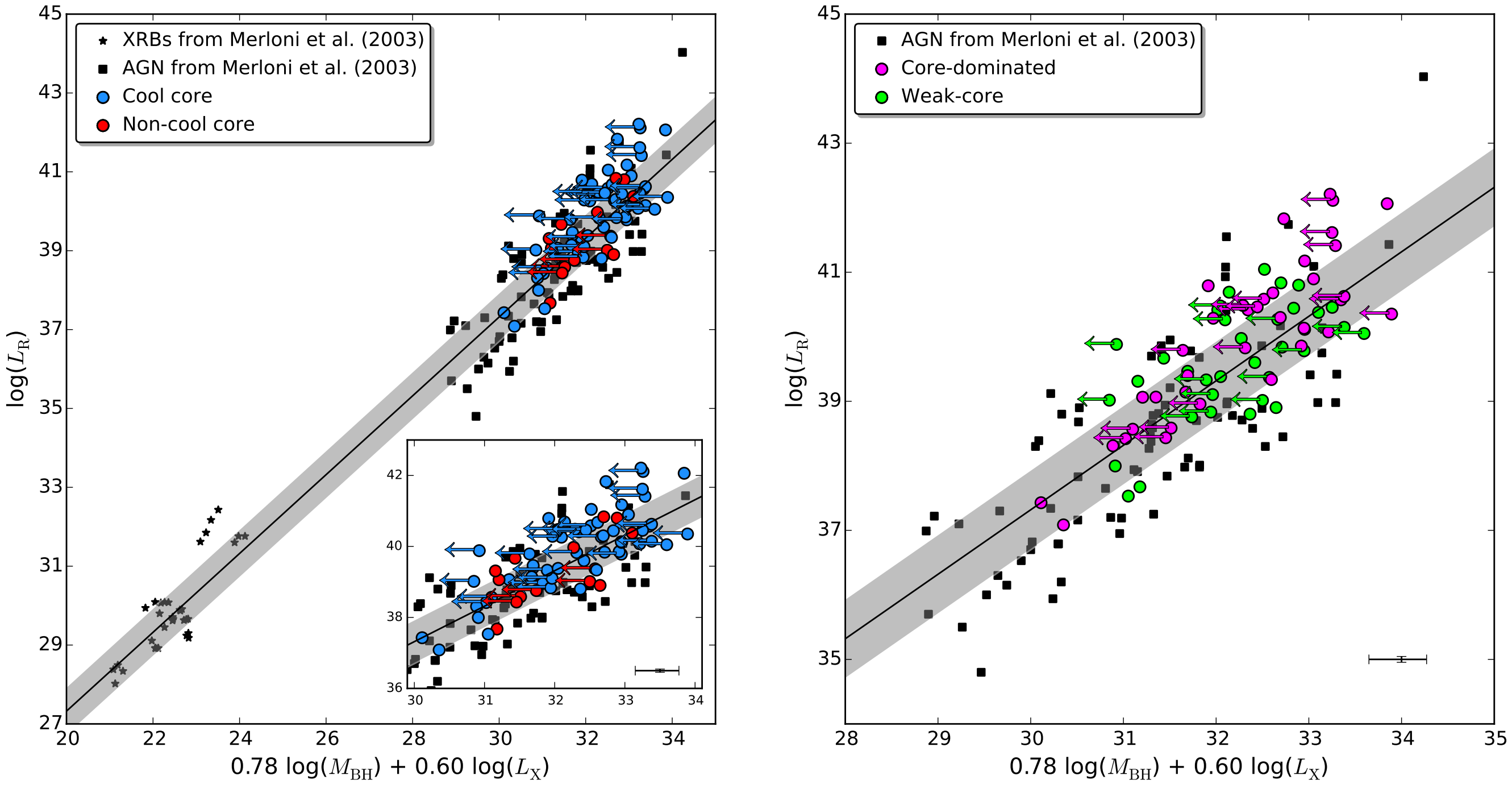

**Figure 2.** Fundamental plane of Merloni et al. (2003). The BH masses are derived from the $M_{\rm BH} - M_{\rm K}$ correlation from Graham & Scott (2013). Upper limits on the X-ray luminosity are shown with arrows. In the left plot cool-cores are shown in blue, non-cool cores in red; in the right plot core-dominated sources are shown in magenta, weak-core sources in green. The XRBs and AGN from Merloni et al. (2003) are shown as black stars and squares, respectively. The scatter of the correlation is shown in grey. A zoom-in of the AGN region is shown on the right panel and in the inset plot on the left panel. The error bars show the mean error of the BH mass, nuclear X-ray luminosity and core radio luminosity.

We find that the horizontal offset visually observed in Figs. 2-3 is positive (above the plane) and significant at $> 2\sigma$ confidence level both for the Merloni et al. (2003) and the Plotkin et al. (2012) correlations, with an average mass offset of 0.3 ± 1.2 and 2.2 ± 1.6, respectively (see Table 1). The offset is also significant for the Körding et al. (2006) correlation (log $\Delta M_{\rm BH} = 2.0 \pm 1.5$, $> 8\sigma$ level) and for the Plotkin et al. (2012) correlation uncorrected from synchrotron cooling (at a $> 7\sigma$ level; see Sect. 4.2), but not for the Gültekin et al. (2009b) correlation (log $\Delta M_{\rm BH} = 0.1 \pm 1.3$ at a confidence level of $> 0\sigma$). In the next sections we repeat the Monte Carlo exercise for different BCG sub-samples (CCs versus NCCs, core-dominated versus weak-core) and discuss some factors that could affect the location of BCGs on the fundamental plane such as synchrotron cooling or the $M_{\rm BH} - M_{\rm K}$ used to estimate the BH mass.

## 4.1 Eddington ratio and Bondi accretion

We derive the Eddington luminosity ($L_{\rm Edd} = 1.3 \times 10^{38} M_{\rm BH}$) and the Eddington ratio ($L_{\rm bol}/L_{\rm Edd}$, where $L_{\rm bol}$ is the bolometric luminosity) for our sample of 72 BCGs. $L_{\rm bol}$ is extrapolated from the spectroscopic $L_{\rm X}$ assuming a bolometric correction factor $k_{\rm bol}$ = 20 (e.g., Vasudevan & Fabian 2007). The distribution of Eddington ratios is plotted in Fig. 4. It ranges from $L_{\rm bol}/L_{\rm Edd} = 5 \times 10^{-8}$ to $L_{\rm bol}/L_{\rm Edd} = 4 \times 10^{-4}$ and peaks at $L_{\rm bol}/L_{\rm Edd} = 10^{-6} - 10^{-5}$. All the 72 BCGs are thus found to be accreting at highly sub-Eddington accretion rates ($L_{\rm bol}/L_{\rm Edd} < 10^{-3}$). We note that the Eddington ratio does not include jet power but is purely dominated by radiative emission. The Eddington-scaled accretion rate of most sources would be typically one order of magnitude higher when including the jet kinetic power to the total emitted luminosity (e.g., Mezcua & Prieto 2014; see Russell et al. 2013 for a study of BCGs that includes cavity power), and thus still be sub-Eddingtonian. Eddington ratios have been calculated using the BH masses derived from the $K$-band magnitudes (eq. 1). If these BH masses were underestimated by the $M_{\rm BH} - M_{\rm K}$ correlation (see Sect. 4.6), the BHs would accrete even more sub-Eddington. Taking for instance log $\Delta M_{\rm BH} = 2.0$, the distribution of Eddington ratios (Fig. 4) would peak at $L_{bol}/L_{Edd} = 10^{-8} - 10^{-7}$ and the tail extend down to $L_{bol}/L_{Edd} = 10^{-10}$. Such low Eddington rates would be of the same order as that of Sgr A*, which has the lowest known BH Eddington ratio ($\sim 10^{-9}$).

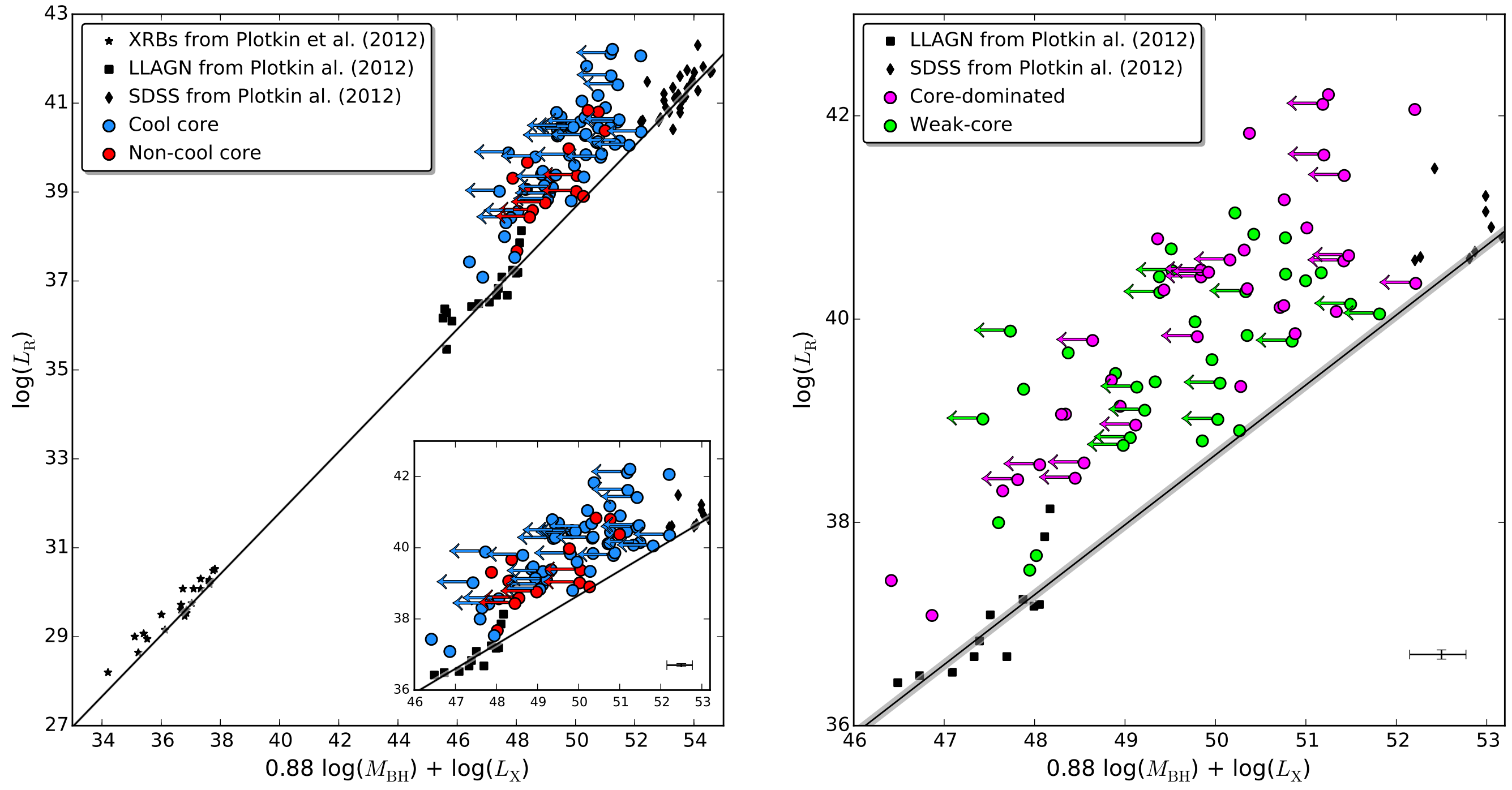

**Figure 3.** Fundamental plane of Plotkin et al. (2012). The BH masses are derived from the $M_{BH} - M_K$ correlation from Graham & Scott (2013). Upper limits on the X-ray luminosity are shown with arrows. In the left plot cool-cores are shown in blue, non-cool cores in red; in the right plot core-dominated sources are shown in magenta, weak-core sources in green. The XRBs from Plotkin et al. (2012) are shown as black stars; the LLAGN and BL Lacs from SDSS as squares and diamonds, respectively. The scatter of the correlation is shown in grey. A zoom-in of the AGN region is shown on the right panel and in the inset plot on the left panel. The error bars show the mean error of the BH mass, nuclear X-ray luminosity and core radio luminosity.

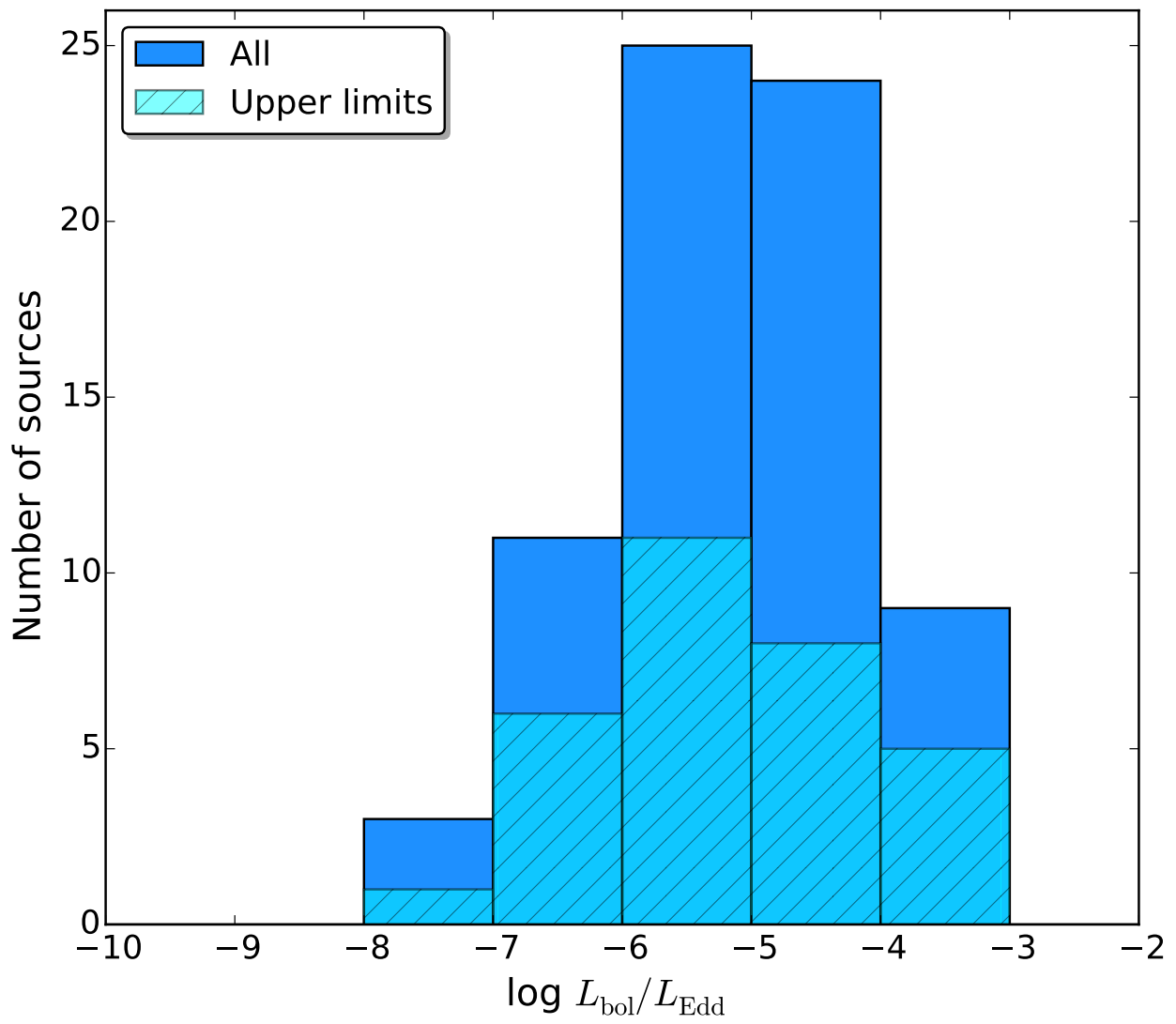

**Figure 4.** Distribution of Eddington ratios for the sample of 72 BCGs (blue bars). Those BCGs with upper limits on the nuclear X-ray luminosity are overploted as hashed bars.

The central short cooling times and peaked X-ray surface brightness profile of CCs indicate that substantial cooling should occur in these clusters. However, only a small percentage of this expected cooling is observed in the form of cold gas and star formation in the local Universe (see e.g., Rawle et al. 2012; McDonald et al. 2012, 2013, 2016; Webb et al. 2015a,b for star-forming BCGs at high redshifts), mainly in relaxed CCs in regions in which, locally, the cooling flow dominates over AGN feedback (see McDonald et al. 2016). CCs have more relaxed and undisturbed X-ray morphologies and, because of their shorter cooling times, their BCGs have a higher gas supply than NCCs, which are dynamically disturbed systems with large cooling times, no central peaked X-ray profiles, and thus with no cooling flow. BCGs in CCs may thus host more massive BHs than those BCGs in NCCs, which should be reflected as CC-BCGs being predominantly more offset from the fundamental plane than NCC-BCGs.

Most of the BCGs in our sample (58 out of 72) are located in CCs. The average offset of these CCs seems to be more significant than that of NCCs for all the fundamental plane correlations considered in Table 1 (e.g., for the Plotkin et al. 2012 fundamental plane, log $\Delta M_{BH} = 2.4 \pm 2.1$ at a $>7\sigma$ confidence level for the CCs and log $\Delta M_{BH} = 2.0 \pm 1.0$ at a $>3\sigma$ confidence level for the NCCs; for the Merloni et al. (2003) fundamental plane, a significant offset at a $>2\sigma$ confidence level is found for the CCs but no significant offset is found for the NCCs). However, this might just be a reflection of the larger scatter of the CC-BCGs around the plane as a significance test does not reject the null hypothesis that there is no difference between the mean BH mass of the AGN in BCGs in CCs and NCCs with 99% confidence both for the Merloni et al. (2003) and the Plotkin et al. (2012) correlations (and also for the other fundamental plane correlations considered in Table 1). To further investigate the different behavior on the fundamental plane of BCGs in CCs and NCCs, we plot the BH mass of their AGN against nuclear X-ray luminosity and core radio luminosity in Fig. 5 (top and bottom, respectively). The BH mass is found to strongly correlate with nuclear (2-10 keV) X-ray luminosity for BCGs located in CCs ($r^2$ = 0.5, p-value = 5 $\times 10^{-9}$) with a slope $m$ = 0.19 ± 0.03; however, no significant correlation is found for those BCGs in NCCs (p-value = 0.1). The BH mass of the AGN in those BCGs in CCs is also observed to correlate with the 5 GHz core radio luminosity ($r^2$ = 0.4, p-value = 1 $\times$ $10^{-7}$) with a slope $m$ = 0.18 ± 0.03, while no significant correlation is observed for those BCGs in NCCs (p-value = 0.2). The above correlations imply that those BCGs with the most powerful (both in terms of nuclear X-ray and radio luminosity) AGN host the most massive BHs, as suggested by Hogan et al. (2015a) to explain their finding that more radio-luminous BCGs tend to inhabit in clusters with higher integrated X-ray luminosity (and thus presumably higher cluster mass and more massive BCGs). However, the finding that X-ray and radio luminosity correlate with BH mass in a significant manner only for those AGN in CC-BCGs suggests that their duty cycle - the time during which the AGN is active- is higher than that of AGN in NCC-BCGs and thus that they might be governed by a different accretion mechanism than that of NCC-BCGs.

Direct Bondi accretion of hot gas could, on average, provide the necessary fuel of AGN in BCGs, specially those with low-power jets (Allen et al. 2006); however, it is not enough to feed the most powerful AGN if these follow the $M_{\rm BH}-\sigma$ relation (e.g., Hardcastle et al. 2007; McNamara et al. 2011; Main et al. 2017). Since the Bondi sphere grows as the square of the BH mass and so the more massive the BH the more hot gas it can accrete, Bondi accretion might still be able to power the most energetic AGN if these are overmassive with respect to the galaxy-BH mass scaling relations. Alternatively, within the cluster core the cooling flow expected in CCs because of their short cooling times can deposit large amounts of cold gas, which can condensate into clumps of cold clouds and infall on to the BH to feed it (e.g., Pizzolato & Soker 2005; Voit & Donahue 2015). The occurrence of such cold accretion in addition to direct Bondi-like hot-phase gas accretion could explain the higher AGN duty cycle of CC-BCGs compared to that of NCC-BCGs, which would be fed only by Bondi accreting gas, and thus the tight correlation between BH mass and AGN output found for the CC-BCGs. The presence of both cold and Bondi-like accretion in the AGN of CC-BCGs was already suggested by Hogan et al. (2015a) to explain the prevalence of core-dominated AGN in CC-BCGs, which the authors do as well associate to a high duty cycle. As we shall see in Sect. 4.4, we also find that most of the BCGs with core-dominated radio emission in our sample (91%) are located in CCs.

Hlavacek-Larrondo et al. (2012b) suggested, based on the lack of a central X-ray point source detection, that strong CCs with X-ray luminosities ≥ $10^{45}$ erg s$^{-1}$ and very energetic AGN feedback might host ultramassive BHs, as these would radiate inefficiently and thus do not present a detectable X-ray nucleus. Our sample includes 10 strong CCs also studied by Hlavacek-Larrondo et al. 2012b (A2204, A1664, RXJ1504.1-0248, A2390, A478, RXJ1720.1+2638, RXJ2129.6+0005, Z3146, Z7160 -MS1455.0+2232 in Hlavacek-Larrondo et al. 2012b-, and A3526 -Centaurus in Hlavacek-Larrondo et al. 2012b-). These strong CCs are not seen to be more offset from the fundamental plane correlations than the CCs nor NCCs; however, the small sample size does not allow us to draw any conclusions.

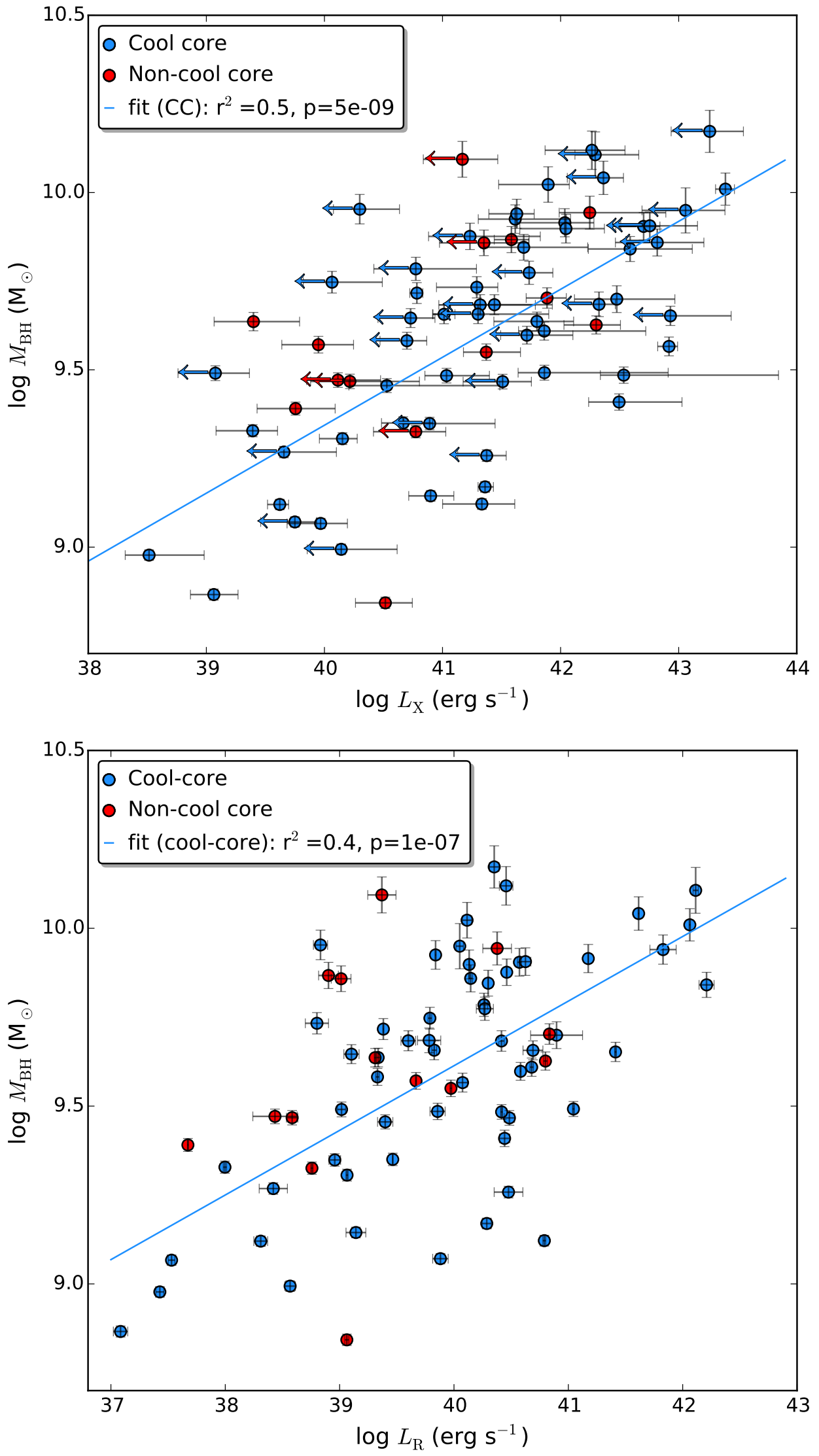

**Figure 5.** BH mass versus 2-10 keV nuclear X-ray luminosity (top) and versus core radio luminosity at 5 GHz (bottom) for the BCGs in cool-core clusters (blue dots) and non-cool core clusters (red dots). Upper limits on the X-ray luminosity are shown as arrows. The solid blue lines show the best-fit linear regressions for the cool-core clusters.

## 4.2 Radiative cooling

The fundamental plane correlations derived assuming a jet model should only be applied if the X-ray emission comes from optically-thin and uncooled synchrotron radiation (e.g., Körding et al. 2006; Plotkin et al. 2012). Because of the broad types of objects and possible diverse origin of the X-ray emission in the Merloni et al. (2003) sample, with their most massive BHs being quasars and thus predominantly with corona-dominated X-ray emission, synchrotron cooling is not corrected for in the Merloni et al. (2003) fundamental plane. The effects of radiative cooling are however taken into account and minimized in the tightest fundamental plane from Plotkin et al. (2012), as they include BL Lacs with jet-dominated X-ray emission.

Synchrotron cooling could strongly affect those SMBHs with $M_{\rm BH}$ > $10^8$ M$_\odot$ (and thus the BCGs here studied) if their X-ray emission comes from the jet, such that the measured X-ray luminosities are lower than the true values expected from the uncooled jet. To correct for this, the 'real' X-ray luminosities should be extrapolated (e.g., via fitting of the SED) from a lower energy band (i.e., the optical) in which the synchrotron emission comes from the uncooled jet and is optically thin (e.g., Körding et al. 2006; Plotkin et al. 2012). While this is feasible for BL Lacs (e.g., Massaro et al. 2004; Plotkin et al. 2012), in BCGs the optical emission of the host galaxy dominates over the AGN emission and thus we did not succeed in obtaining an optical-extrapolated jet X-ray luminosity via SED fitting. Plotkin et al. (2012) provide the fundamental plane coefficients that would be obtained when not correcting for synchrotron cooling, this is, when using the real X-ray luminosities, and find that the slopes are shallower than expected from an optically-thin jet. If BCGs were strongly affected by radiative cooling, they would be expected to lie closer to this shallower fundamental plane. Nonetheless, using this cooling-uncorrected correlation we still find that BCGs are on average offset by log $\Delta M_{BH}$ = 2.1 ± 2.6 and that this offset is significant at a $>7\sigma$ level (see Table 1). A significance test does not reject either (p-value = 0.66) the null hypothesis that the offset from the Plotkin et al. (2012) correlation with and without cooling correction is different. Even if the X-ray luminosities were higher by an extreme factor 12 (probed by Plotkin et al. (2012) by simulating that X-ray satellites are able to observe optically-thin synchrotron emission from the jet of a low-mass BH of $< 10^8$ $M_\odot$ but X-ray emission from the synchrotron-cooled jet of a higher mass BH), they would still be offset from the Plotkin et al. (2012) correlation (cooling corrected) at $>2\sigma$. The BCGs would be located on the fundamental plane if their nuclear X-ray luminosities were a factor 100 higher; however, in that case the Eddington ratios and Eddington-scaled accretion rates would be one order of magnitude higher than the typical values for BCGs (e.g., Russell et al. 2013). Synchrotron cooling seems thus not to explain the observed offset of BCGs from the fundamental plane.

### 4.3 Emission processes

The radio/X-ray term (log $L_R$ ~ 0.7log $L_X$) of the above fundamental plane relations stems from a correlation originally found for low/hard and quiescent state XRBs (Corbel et al. 2003; Gallo et al. 2003, 2006). Yuan & Cui (2005) argue that the coefficient $\xi_X$ ~ 0.7 holds only for sources accreting above a critical X-ray luminosity $L_{X,\mathrm{critical}} \sim 10^{-5} - 10^{-6} L_{\mathrm{Edd}}$ and whose X-ray emission is dominated by the accretion flow, while for sources with very low accretion ($L_X < L_{X,\mathrm{critical}}$) the X-ray emission is dominated by the jet and the coefficient should steepen to log $L_R$ ~ 1.2-1.3 log $L_X$. This prediction was proven by Yuan et al. (2009) and Xie & Yuan (2017): using a sample of 22 and 75 quiescent AGN, respectively, with $L_X/L_{\mathrm{Edd}} \lesssim 10^{-6}$, they found a coefficient in the log $L_R/L_X$ term of the fundamental plane of $\xi_X$ ~1.22-1.23. Using a sample of 16 LLAGN, de Gasperin et al. (2011) find that their coefficients in the fundamental plane relation are also consistent with those of Yuan et al. (2009). Xie & Yuan (2017) additionally explore the radio/X-ray correlation using only $L_R$ and $L_X$ (i.e. using a $M_{BH}$ free sample). This allows them to derive a more direct constrain to the coefficient of $\xi_X$ ~1.36.

To probe whether the steeper slope, and thus different fundamental plane, found by Yuan et al. (2009) and Xie & Yuan (2017) for quiescent sources could explain the offset of BCGs from the fundamental plane, we should divide our sources into quiescent (i.e. with $L_X/L_{\mathrm{Edd}} < 10^{-6}$) and non-quiescent and plot their $L_R$ versus $L_X$. We find that most of the BCGs are quiescent (44 sources) and 28 non-quiescent. However, these non-quiescent BCGs are simply the most X-ray luminous ones: all the BCGs with $L_X > 10^{42}$ erg s$^{-1}$ are found to be non-quiescent. Given the scatter in $L_R$ versus $L_X$ for $L_X > 10^{42}$ erg s$^{-1}$ (see Fig. 6, bottom), this subset shows no significant correlation (p-value = 0.4) and thus the change in slope predicted by Yuan & Cui (2005) cannot be tested. The fit of a linear regression to $L_R$ versus $L_X$ for the full BCG sample yields a slope $m$ = 0.75 ± 0.08 ($r^2$ = 0.6) and a probability for rejecting the null hypothesis that there is no correlation of $7 \times 10^{-15}$ (> 99.9% confidence). The same slope ($m$ = 0.8 ± 0.1) is found for the sub-sample of quiescent BCGs (with $r^2$ = 0.5, p-value = $1 \times 10^{-7}$). We note that this slope could be flatter given that some of the nuclear X-ray luminosities are upper limits. The quiescent BCGs (and the full sample of BCGs) do thus not follow the steep slope of $\xi_X$ ~1.36 found by Xie & Yuan (2017) using a $M_{BH}$ free sample, which can thus not explain the on average offset of the BCGs from the fundamental plane.

All the BCGs in our sample host BHs accreting at sub-Eddington rates, which are thought to be fed by radiatively inefficient accretion flows (RIAFs). In such BHs both optically-thin jet synchrotron emission and inverse Compton off an X-ray corona emission are present. The fundamental planes coefficients found by Plotkin et al. (2012), Körding et al. (2006) and Xie & Yuan (2017) favor inverse Compton over jet synchrotron as the dominant mechanism of X-ray emission, while the Merloni et al. (2003) regression favors a RIAF model with inverse Compton X-ray emission. The specific RIAF model considered by Merloni et al. (2003) is an advection-dominated accretion flow (ADAF) with a radiative efficiency of the accretion flow $q$ = 2.3 ($q$ = 1 for radiatively efficient accretion). However, there exist alternative models to the ADAF (e.g., convection-dominated accretion flows or CDAFs, Narayan et al. 2000; advection-dominated inflow-outflow solutions or ADIOS, Blandford & Begelman 1999) in which thermal bremsstrahlung with $q$ = 2 is the most likely dominant mechanism of X-ray emission.

In ADAF models with accretion rates $< 10^{-3}$, the radio luminosity is predicted to scale with X-ray luminosity as $L_R \propto L_X^{0.6}$ (e.g., Yi & Boughn 1998). The finding that $L_X$ correlates with $L_R$ with a slope $m$ = 0.75 ± 0.08 suggests that the *X-ray emission originates predominantly from an ADAF* with a small contribution from jet synchrotron emission. This might explain why the offsets from the fundamental plane are larger for the Plotkin et al. (2012) and Körding et al. (2006) correlations (which favor the jet model) than for the Merloni et al. (2003) correlation (which favors an ADAF model; see Table 1), and also why synchrotron cooling (which affects the X-ray emission only if it originates from a jet) cannot account for the observed offsets (see Sect. 4.2). Modeling of the nuclear SED would be required to further disentangle the ADAF and jet contribution to the origin of the X-ray emission. This was performed by Wu et al. (2007) for one of the sources in our sample, A2052 (3C 317), finding that the hard X-ray spectrum can be well fitted by the sum of a jet and ADAF model. Russell et al. (2013) argue that the linear trend between nuclear X-ray luminosity and 5 GHz core radio luminosity they find for a sample of 22 BCGs is likely a distance effect based on their lack of correlation between the X-ray and radio flux. By studying a larger sample of 72 BCGs, we find that the nuclear X-ray flux and 5 GHz radio core flux correlate at a significant level (> 99.9% confidence) with a slope $m$ = 0.75 ± 0.08 (Fig. 6). As we will see later, this correlation is stronger when considering only core-dominated BCGs (Sect. 4.4).

The fundamental plane coefficients (of both the corona and the jet models) depend on the radiative efficiency of the accretion flow $q$,

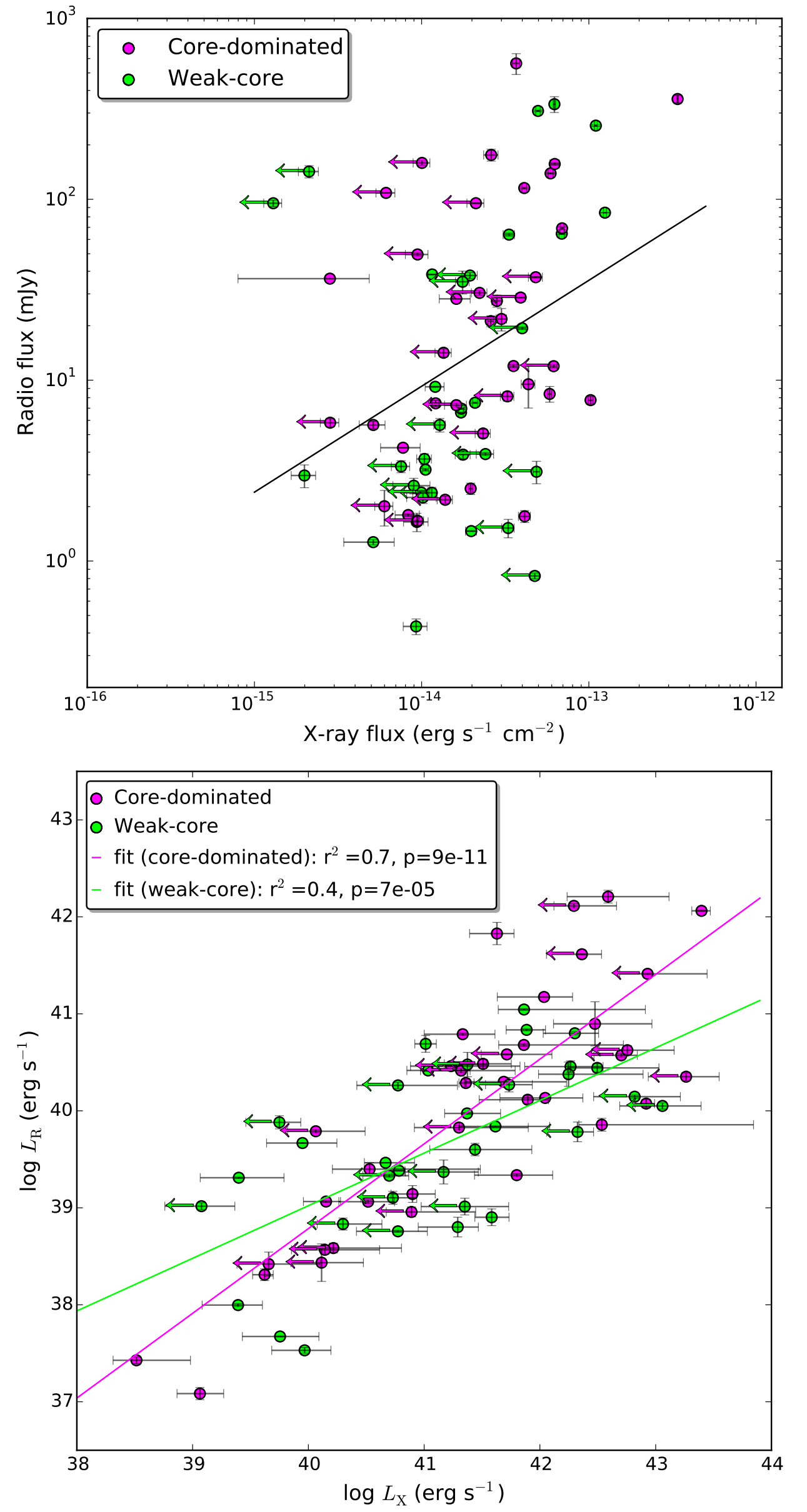

**Figure 6.** *Top*: 5 GHz core radio flux versus nuclear X-ray flux in the 2-10 keV band for the core-dominated BCGs (magenta dots) and weak-core BCGs (lime dots). Upper limits on the X-ray flux are shown as arrows. The black solid line shows the slope of the linear regression for the full sample of 72 BCGs, $m$ = 0.6 ± 0.2, p-value = 0.003. *Bottom*: Core radio luminosity at 5 GHz versus nuclear X-ray luminosity (2-10 keV) for the core-dominated BCGs (magenta dots) and weak-core BCGs (lime dots). Upper limits on the X-ray luminosity are shown with arrows. The solid lines show the best-fit linear regressions for the core-dominated (magenta line) and weak-core (lime line) BCGs.

on the power-law index $p$ of the relativistic electrons accelerated in the jet ($p$ = $2\alpha_X$ + 1 for optically-thin synchrotron emission, where $\alpha_X$ is the observed X-ray spectral index and has typical values $\alpha_X$ = 0.5-1), and on the radio spectral index $\alpha_R$. Different accretion processes imply as well different values of the magnetic field at the base of the jet as a function of BH mass ($\frac{\partial \ln \Phi_B}{\partial \ln M}$) and accretion rate ($\frac{\partial \ln \Phi_B}{\partial \ln m}$); however, for most RIAF models $\frac{\partial \ln \Phi_B}{\partial \ln M}$ = -1/2 and $\frac{\partial \ln \Phi_B}{\partial \ln m}$ = 1/2 (see Table 3 in Merloni et al. 2003). We thus consider different values of $q$, $\alpha_X$ and $\alpha_R$ to probe whether the different predicted fundamental plane coefficients can put the BCGs on average on the fundamental plane correlation.

Plotkin et al. (2012) find that increasing (decreasing) $\alpha_X$ to 0.7 (0.5), instead of using the $\alpha_X$ = 0.6 assumed to extrapolate X-ray luminosities from optical nuclear luminosities, would increase (decrease) the X-ray luminosities by a factor 2. While this would bias the fundamental plane towards flatter (steeper) slopes, the coefficients would still be consistent within $3\sigma$ with those found assuming $\alpha_X$ = 0.6 (see Fig. 3 in Plotkin et al. 2012) and would thus not significantly change the offset of BCGs from the fundamental plane. Note that an $\alpha_X$ = 0.5 ($p$ = 2) is also considered by Merloni et al. (2003), for which we find that the BCGs are significantly offset at $>2\sigma$ confidence level (Table 1). The most extreme case would be that of $\alpha_X$ = 1, which would push the coefficients to even shallower slopes and be equivalent to modifying the X-ray luminosities by a factor 12 (Plotkin et al. 2012). However, as we saw in Sect. 4.2, using the cooling-uncorrected fundamental plane of Plotkin et al. (2012) with shallower slopes does not put them on average on the fundamental plane; they would still be offset from it at a $>7\sigma$ level confidence. We also test whether an ADAF (instead of a jet) model with $q$ =2 and $q$ = 2.3 is able to put the BCGs on average on the fundamental plane of Plotkin et al. (2012). We find that the BCGs keep being on average positively offset from the plane by log $\Delta M_{BH}$ = 1.9 ± 1.2 ($q$ =2) and log $\Delta M_{BH}$ = 6.1 ± 1.1 ($q$ =2.3) at $>8\sigma$.

The radio luminosities of the 72 BCGs are attributed to a core component and have been derived considering typically a flat or inverted radio spectral index ($\alpha_R$ < 0.5, where $S \propto \nu^{-\alpha_R}$; Hogan et al. 2015a) in agreement with the canonical $\alpha_R$ assumed by Merloni et al. (2003) (flat, $\alpha_R$ = 0) and Plotkin et al. (2012) (inverted, $\alpha_R$ = -0.15) in their best-fit fundamental plane correlations. Different values of $\alpha_R$ predict different fundamental plane coefficients (see Fig. 4 in Merloni et al. 2003 and Fig. 1 in Plotkin et al. 2012) and thus we could in principle investigate the modifications induced on the fundamental plane coefficients by $\alpha_R$. However, for this we should assume that the core components of all the BCGs are characterized by the same spectral index, while Hogan et al. (2015a) illustrate a diversity in spectral index for a large sample of cores (from $\alpha_R$ = -1 to $\alpha_R$ = 0.7). It is thus meaningless to perform this analysis by assuming that the spectral index is a constant, as the core components have spectral turnovers in the 0.5-50 GHz range (Hogan et al. 2015a). The only way to get round this is to use a higher frequency than 5 GHz (e.g., at 150 GHz), as we do in the next section.

### 4.4 Core-dominated BCGs

The tightest fundamental plane correlations are found for sub-Eddington accreting sources with flat or inverted radio spectra and thus whose radio emission comes from the core of the jet (i.e. optically-thick; e.g., Körding et al. 2006; Plotkin et al. 2012), while a broader scatter is obtained when the radio luminosity is a mixture of both core and extended jet emission (e.g., the Merloni et al. 2003 correlation). In the right panel of Fig. 2 we locate the 72 BCGs on the fundamental plane of Merloni et al. (2003) distinguishing between core-dominated sources (magenta dots) and sources with significant extended radio emission in addition to the weak core (weak-core BCGs; green dots).

While weak-core BCGs tend to be equally distributed above, on, and below the fundamental plane of Merloni et al. (2003), those core-dominated BCGs seem to show a higher trend for lying on and above the correlation (with only one core-dominated BCGs lying below it). When considering the Plotkin et al. (2012) correlation, the only BCG that sits on the fundamental plane is a weak-core; all the others (both core-dominated and weak-core BCGs) sit above the fundamental plane.

We further check whether there are any differences between core-dominated and weak-core BCGs by plotting their radio luminosity versus BH mass. These two variables were found to be correlated for samples of SMBHs only (e.g., Franceschini et al. 1998; Nagar et al. 2002) and for the full sample of XRBs and SMBHs of Merloni et al. (2003), leading to the discovery of a mass segregation and (together with the dependence of $L_R$ also on X-ray luminosity; e.g., Merloni et al. 2003; Falcke et al. 2004) consequent fundamental plane of BH accretion. We find that the 5 GHz core radio luminosity strongly correlates with BH mass for the core-dominated BCGs ($r^2$ = 0.6, p-value = $6 \times 10^{-8}$, see Fig. 7), but no significant correlation is observed for the weak-core sources (p-value = 0.1, $r^2$ = 0.07). The 5 GHz core radio luminosity is also more strongly correlated with nuclear X-ray luminosity for the core-dominated BCGs ($r^2$ = 0.7, p-value = $9 \times 10^{-11}$) than for the weak-core sources ($r^2$ = 0.4, p-value = $7 \times 10^{-5}$ see Fig. 6 bottom).

The correlations between $L_R$ and $M_{BH}$, and $L_R$ and $L_X$ are expected if the jet physics is scale invariant, and were already found in the fundamental plane discovery papers (e.g., Heinz & Sunyaev 2003; Merloni et al. 2003; Falcke et al. 2004; see also e.g., Fabbiano et al. 1989; Yi & Boughn 1998; Canosa et al. 1999). We find that these correlations are more significant for the core-dominated BCGs, even if the 5 GHz core radio luminosity is always plotted (both for the core-dominated and weak-core BCGs), which suggests that only core-dominated sources should be included in the fundamental plane analysis. This was already proven by Merloni et al. (2003) by finding that the scatter of their correlation is significantly reduced when considering a sub-sample of only flat-spectrum sources. For those sources with a flat radio spectral index, we can be confident that their radio emission comes from the core of the jet, which is a pre-requisite of scale-invariant jet models (Heinz & Sunyaev 2003). As we will see later, even restricting our sample of BCGs to strongly core-dominated radio sources, BCGs are still significantly offset from the fundamental plane correlations.

Of the 38 BCGs with core-dominated radio emission, 35 are located in CCs and in which there might be more gas available for accretion (compared to NCCs where no gas cooling is expected to occur because of the larger cooling time). The pre-eminence of the core radio emission indicates that the AGN in core-dominated BCGs are currently active and maybe accreting at a significant rate (Hogan et al. 2015a,b). To test this we check whether there is any correlation between radio luminosity or BH mass and Eddington ratio and whether this correlation is stronger for the core-dominated than for the weak-core BCGs. The BH mass and Eddington ratio are found to significantly correlate for core-dominated BCGs (p-value = $4 \times 10^{-4}$, $r^2$ = 0.3) but not for those BCGs with weak core radio emission (p-value = 0.1). The core radio luminosity and Eddington ratio are also found to more strongly correlate for core-dominated BCGs (p-value = $4 \times 10^{-8}$, $r^2$ = 0.6) than for weak-core BCGs (p-value = $8 \times 10^{-5}$, $r^2$ = 0.4; see Fig. 8). We also find that the core-dominated BCGs have a median Eddington ratio a factor ~3 higher than that of the weak-core sources, further supporting that the AGN in the core-dominated BCGs are currently more active than those in the weak-core BCGs.

Among the sample of 72 BCGs, there are 14 with radio detections at 150 GHz, i.e., for which the radio emission comes mostly from the core (Hogan et al. 2015b). They are the most core-dominated sources in our sample and they all reside in CC clusters (Hogan et al. 2015b). In what follows, we will use this sub-sample of BCGs with 150 GHz radio emission to further probe the differences so far observed for the core-dominated BCGs and their offset from the fundamental plane.

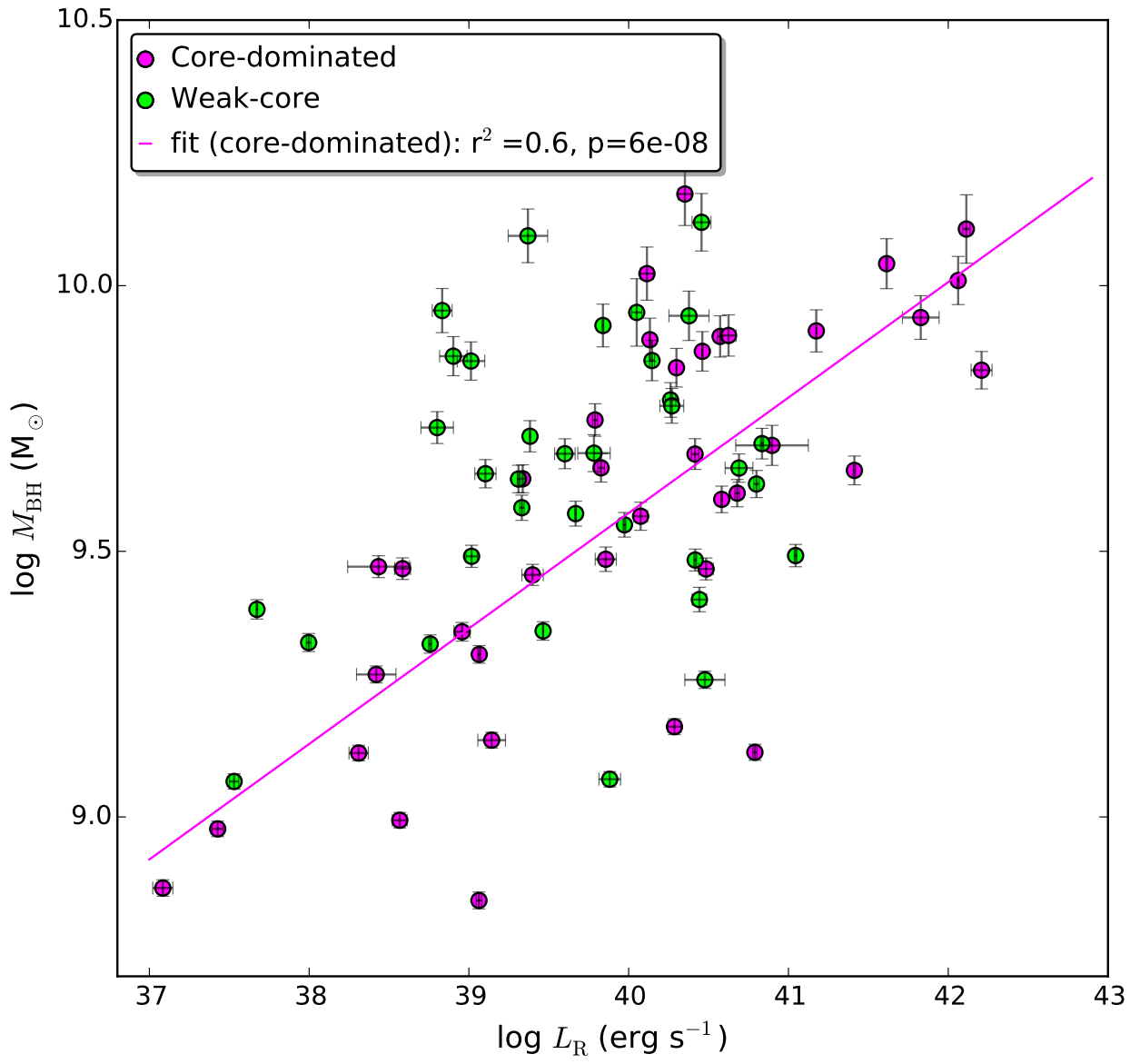

**Figure 7.** BH mass versus core radio luminosity at 5 GHz for the core-dominated (magenta dots) and weak-core (lime dots) BCGs. The solid line show the best-fit linear regression for the core-dominated sources (magenta line).

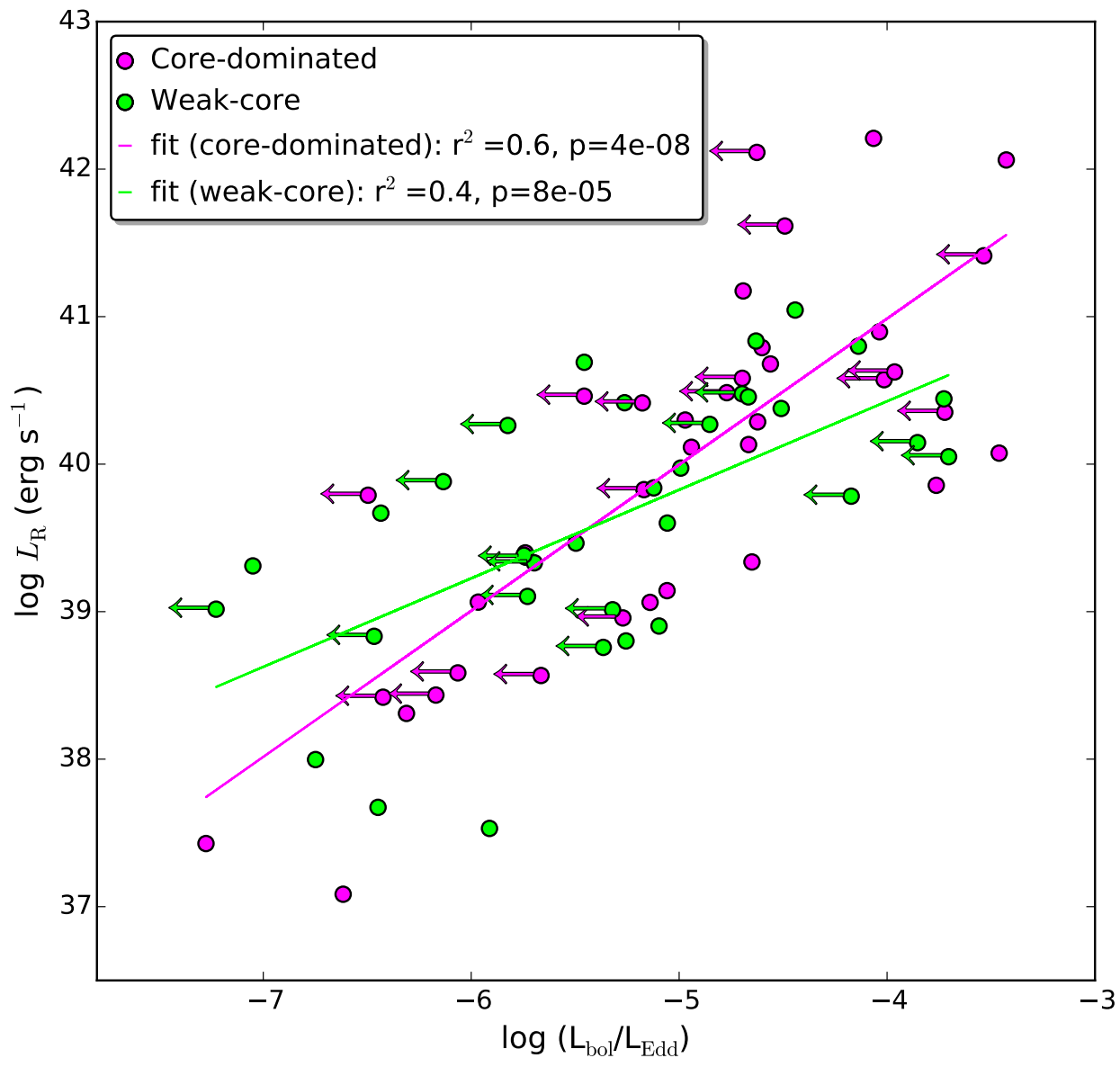

**Figure 8.** Core radio luminosity at 5 GHz versus Eddington ratio for the core-dominated BCGs (magenta dots) and weak-core BCGs (lime dots). Upper limits on the X-ray luminosity are shown with arrows. The solid lines show the best-fit linear regressions for the core-dominated (magenta line) and weak-core (lime line).

In Fig. 9 we locate the 14 sources with radio emission at 150 GHz on the fundamental planes of Merloni et al. (2003) (left) and Plotkin et al. (2012) (right) using the BH masses derived from $M_K$ and the 5 GHz radio luminosities. We see that all the sources lie above the correlations, with an average mass offset of log $\Delta M_{BH}$

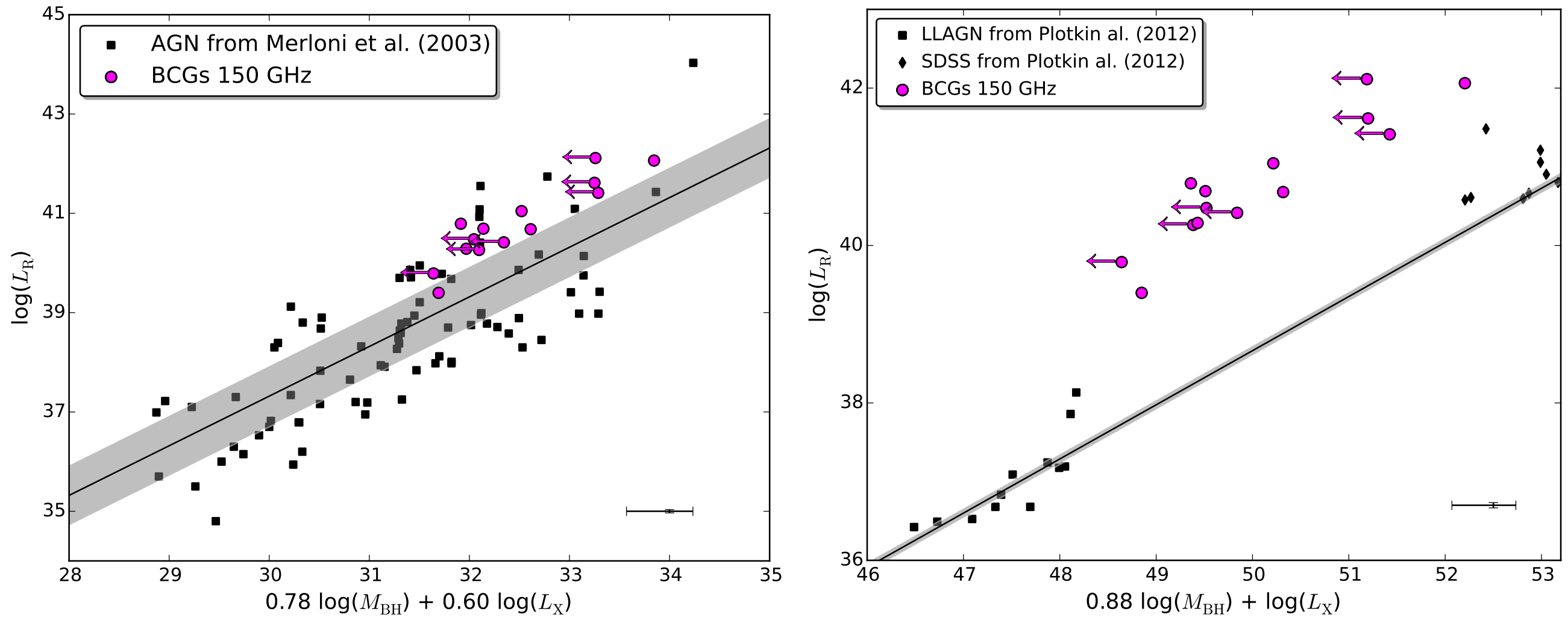

**Figure 9.** Zoom-in of the AGN region of the fundamental plane of Merloni et al. (2003) (left) and of Plotkin et al. (2012) (right) for the BCGs with 150 GHz radio emission. The BH masses are derived from the $M_{BH} - M_K$ correlation from Graham & Scott (2013). Upper limits on the X-ray luminosity are shown with arrows. The AGN from Merloni et al. (2003) are shown as black squares (left panel). The LLAGN and SDSS BL Lacs from Plotkin et al. (2012) are shown as squares and diamonds, respectively (right panel). The error bars show the mean error of the BH mass, nuclear X-ray luminosity and core radio luminosity.

= 1.3 ± 1.0 (probability that the offset is significant >3$\sigma$) and log $\Delta M_{BH}$ = 3.4 ± 1.4 (>5$\sigma$) for the Merloni et al. (2003) and Plotkin et al. (2012) correlations, respectively. The sources are also significantly offset for the other fundamental plane correlations considered in Table 1. We test whether these offsets are significantly different that those found for the full sample of BCGs. We find that the null hypothesis (that there is no difference between the average mass offsets of the full sample and the 150 GHz sub-sample) can be rejected with 99.9% confidence for the Merloni et al. (2003), Körding et al. (2006), Plotkin et al. (2012) and Gültekin et al. (2009b) fundamental plane correlations. From all the above we can thus conclude that *the offset of BCGs from the fundamental plane is more significant for core-dominated BCGs.*

### 4.5 $M_{BH} - M_K$ relation

The BH masses have been calculated from the $K$-band magnitudes using the Graham & Scott (2013) correlation for spheroids (see Eq. 1), which includes core-Sérsic galaxies, has a scatter of 0.44 dex, and is corrected for the underestimation of the $K$-band magnitudes provided by the 2MASS XSC (Graham & Scott 2013). Using the on average 0.3 mag fainter magnitudes *k_m_ext* tabulated in the XSC the mass offset of the BCGs from the fundamental plane would be 1$\sigma$ more significant. The use of a $M_{BH} - M_K$ relation that takes into account the 2MASS XSC underestimated magnitudes is thus required in order to study the location of the BCGs on the fundamental plane. Another recent $M_{BH} - M_K$ correlation that corrects for the 2MASS offset is that of Kormendy & Ho (2013) (see their equation 2), which has an intrinsic scatter of 0.31 dex and includes elliptical galaxies and classical bulges. We find that the BH masses derived using this correlation are consistent within the errors with those derived from the Graham & Scott (2013) relation for those BCGs with $M_K \geq -25.5$ mag. However, for those sources with $M_K \lesssim 26$ (or high $M_{BH}$) the BH mass derived from Kormendy & Ho (2013) is ~2 times higher than that from Graham & Scott (2013) (see Fig. 10). This difference can be attributed to the different slopes of the $M_{BH} - M_K$ correlation, which Graham & Scott (2013) found to bend at high $M_{BH}$ (i.e., for $M_K$ < 25; see figure 3 in Graham & Scott 2013; see also Krajnović et al. 2017 for a bend in the $M_{BH} - M_*$ correlation) towards a lower value than that reported by Kormendy & Ho (2013). The exclusion in the sample of Kormendy & Ho (2013) of sources with BH masses based on ionized gas rotation curves and the inclusion in the sample of Graham & Scott (2013) of some sources that could have underestimated BH masses seems to be the explanation behind the discrepant slopes (see Kormendy & Ho 2013 for a detailed discussion). The use of the Kormendy & Ho (2013) $M_{BH} - M_K$ correlation to derive the BH masses of our BCGs decreases the significance of their horizontal average offset from the fundamental plane by 1-2$\sigma$; yet, the BCGs are found to be still on average positively offset from the fundamental plane correlations of Merloni et al. (2003), Plotkin et al. (2012), and Körding et al. (2006) by as much as >8$\sigma$.

When considering the Graham (2007) correlation used by Hlavacek-Larrondo et al. (2012b) (which was the latest one available at that time) we find that the BH masses are on average a factor ~3 lower than those obtained from the Graham & Scott (2013) relation, so that using the Graham (2007) relation the BCGs would move leftwards and lie even more offset (above) the fundamental plane. These differences can be attributed to the sample used by Graham (2007), whose correlation is based on a sample of elliptical and disc galaxies while BCGs are hosted by massive elliptical galaxies. The relations from Graham & Scott (2013) or Kormendy & Ho (2013) are thus more appropriate for BCGs.

### 4.6 Ultramassive BHs in BCGs

If, as discussed in the previous sections, the accretion physics governing BCGs is not different from that of stellar-mass/SMBHs because of e.g., BCGs being surrounded by a large amount of hot gas, and their radio and X-ray luminosities have not been over/underestimated, then BCGs should sit on average on the fundamental plane of BH accretion. The finding that they are positively offset from it may imply that their BH masses are underestimated by the $M_{BH} - M_K$ relation, this is, that BCGs are overmassive with respect to the BH mass scaling relations (see Fig. 11). This was already suggested for a few BCGs for which dynamical modeling revealed BH masses larger than predicted from the $M_{\rm BH} - \sigma$ and $M_{\rm BH} - M_{\rm K}$ relations (e.g., M87, Gebhardt et al. 2011; the BCGs in A1836 and A3565, Dalla Bontà et al. 2009; NGC 3842 in A1367, McConnell et al. 2011a), with the most extreme cases being that of the BCG in Abell 1201 (Smith et al. 2017a,b) or NGC 4889 (in Coma; McConnell et al. 2011a) for which a BH mass of more than $10^{10}$ $\mathrm{M_\odot}$ was measured.

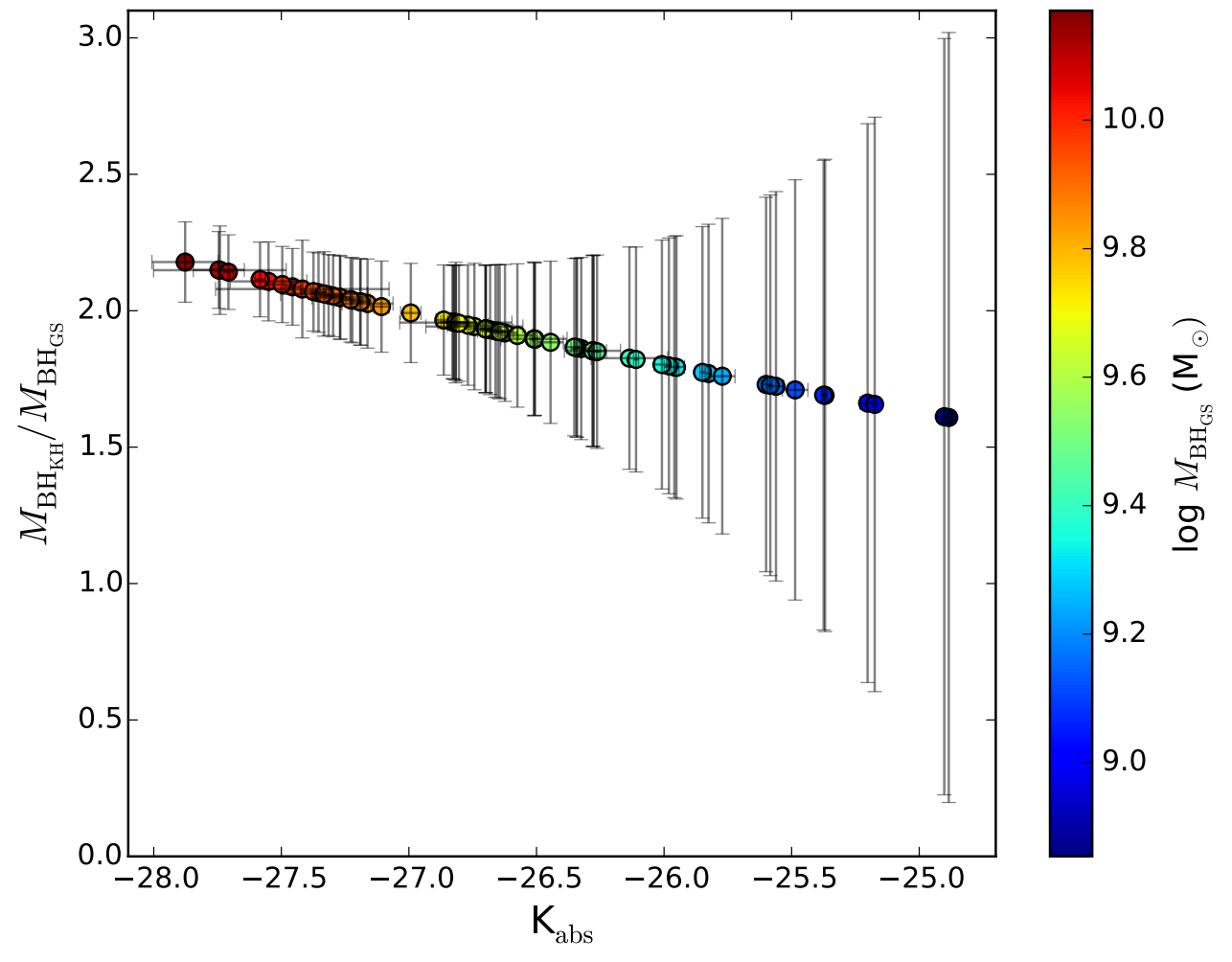

**Figure 10.** Ratio of the BH mass derived from the $M_{\rm BH} - M_{\rm K}$ relation from Kormendy & Ho (2013) ($M_{\rm BH_{KH}}$) and that derived from Graham & Scott (2013) ($M_{\rm BH_{GS}}$) versus $K$-band absolute magnitude. The BH mass derived from Graham & Scott (2013) is shown as a color bar.

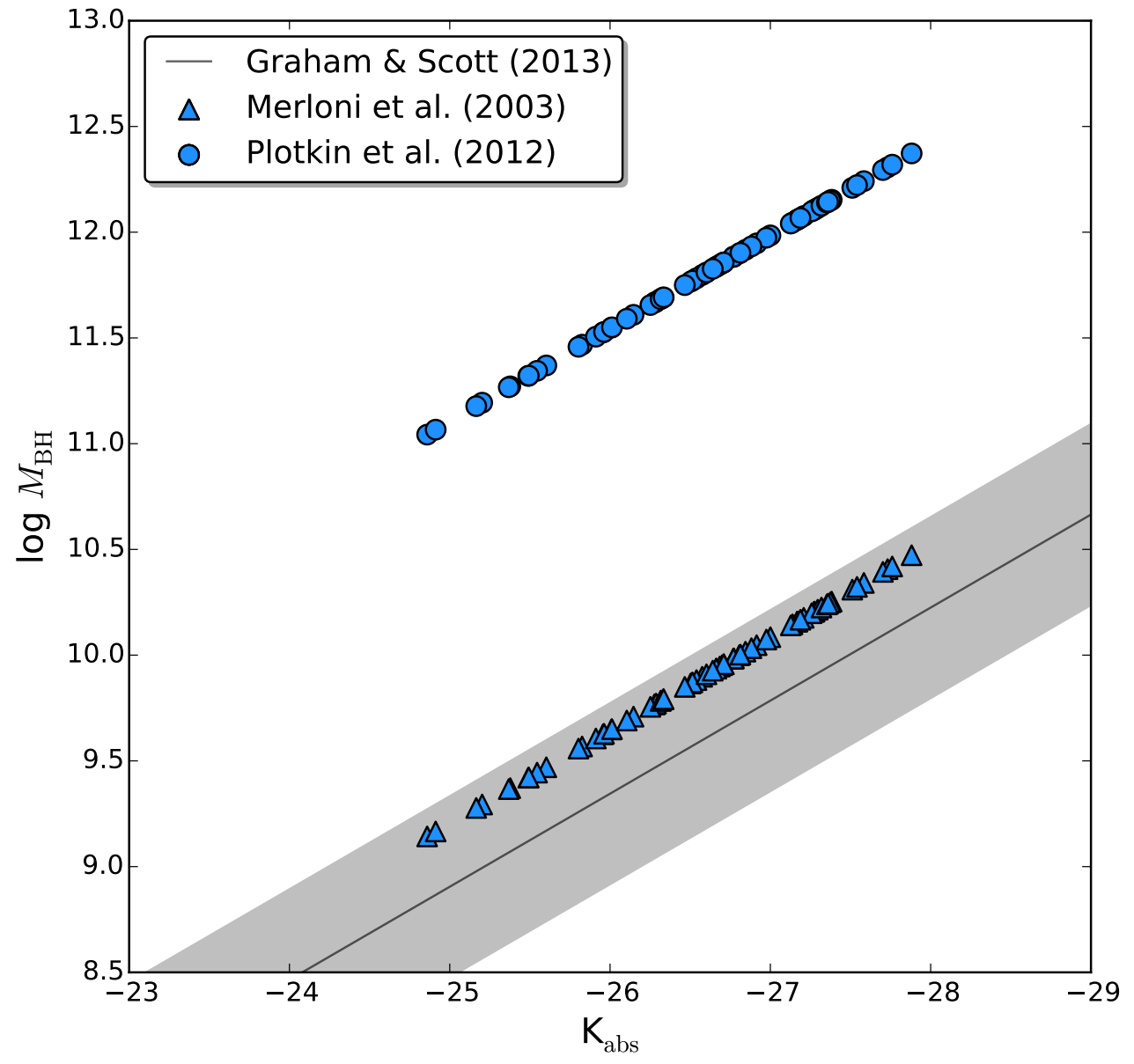

**Figure 11.** BH mass versus $K$-band luminosity for the sample of 72 BCGs using the average BH mass offsets estimated from the fundamental plane of Merloni et al. (2013, triangles) and of Plotkin et al. (2012, circles). The $M_{\rm BH} - M_{\rm K}$ relation from Graham & Scott (2013) is shown as a grey line. The scatter of the correlation (of 0.44 dex) is shown as a shaded area.

The average mass offsets needed for BCGs to sit on the fundamental plane are reported in Table 1 and are as high as log $\Delta \mathrm{M_{BH}} \sim 2$ when considering the full sample of 72 BCGs and the fundamental planes of Plotkin et al. (2012) and Körding et al. (2006). However, the nuclear X-ray emission of BCGs seems to come predominantly from an ADAF (see Sect. 4.3), as favored by the Merloni et al. (2003) correlation but not by the other fundamental planes reported in Table 1 (e.g., Körding et al. 2006; Plotkin et al. 2012) which favor the jet model. Therefore, and to be as conservative as possible, we derive the BH masses implied from the fundamental plane (i.e. those for which BCGs sit, on average, on the fundamental plane) by considering the mass offset obtained from the Merloni et al. (2003) correlation for the full sample of 72 BCGs (log $\Delta \mathrm{M_{BH}} = 0.3$). We note that, despite of the increase in sample size and detailed analysis (which includes now nuclear X-ray detections and NCCs), this offset is consistent with that found by Hlavacek-Larrondo et al. (2012b) for a sample of 18 BCGs with no detectable X-ray nucleus and residing in strong CCs. Considering a minimum average offset of 0.3, we find that the BCG BH masses derived from the fundamental plane lie between $1\times10^9$ and $3\times10^{10}$ $\mathrm{M_\odot}$ and that *~40% of the BCGs (28 out of 72) are ultramassive with $M_{\rm BH} > 10^{10}$ $M_\odot$*. The distribution of these BH masses is plotted in Fig. 12, where we show for comparison the BH mass distribution that would be obtained considering the average mass offset of the Plotkin et al. (2012) correlation. While the BH mass distribution derived from the correlation of Merloni et al. (2003) peaks at log $M_{\rm BH} \sim 10^{10}$ $\mathrm{M_\odot}$, that from Plotkin et al. (2012) does it at log $M_{\rm BH} \sim 10^{12}$ $\mathrm{M_\odot}$. Assuming $\sigma \sim 300$ km s$^{-1}$ (e.g., Dalla Bontà et al. 2009; McConnell et al. 2012), the sphere of influence $r_{\rm inf}$ (kpc) = 4.302 $\times10^{-6} M_{\rm BH}/\sigma^2$ of a $10^{10}$ $\mathrm{M_\odot}$ BH would be of ~ 0.5 kpc, consistent with observations of the gas and stellar kinematics of BCGs (e.g., McConnell et al. 2012). However, a $10^{12}$ $\mathrm{M_\odot}$ BH would have an unrealistic $r_{\rm inf}$ of ~ 50 kpc, which reinforces our previous suggestion that the use of a fundamental plane correlation that favors the jet model (i.e. that of Plotkin et al. 2012 or Körding et al. 2006) is not the most appropriate for estimating BH masses in BCGs.

The presence of ultramassive BHs in the cores of galaxy clusters would be expected if SMBHs correlate with their host dark matter halos (e.g., Kormendy & Bender 2011). The existence of a $M_{\rm BH} - V_{\rm c}$ (where $V_{\rm c}$ is the circular velocity of the host galaxies and is taken as a proxy for dark matter halo mass) correlation has been long debated and, if any (Kormendy & Bender 2011), it seems to be strongly dependent on BH mass measurement and galaxy type (e.g., Ferrarese 2002; Baes et al. 2003; Ho 2007; Volonteri et al. 2011; Beifiori et al. 2012; Kormendy & Ho 2013; Sabra et al. 2015). While a large scatter is found for low $V_{\rm c}$ and low BH masses (i.e. for spiral galaxies; e.g., Sabra et al. 2015), the relation seems to be tighter for classical bulges and ellipticals, where $V_{\rm c}$ is however not well known (Kormendy & Ho 2013). The presence of ultramassive BHs in BCGs would thus support the coupling of dark matter halos and central BHs for BCGs with bulges, in agreement with the recent relationship between jet power (assumed to scale with BH mass; e.g., Somerville et al. 2008) and halo mass found for a sample of 45 BCGs with cooling times shorter than 1 Gyr (Main et al. 2017).

Several types of sources in addition to BCGs have been found to be overmassive with respect to their bulge luminosity or mass (i.e. $M_{\rm BH}/M_{\rm bulge} > 5\%$, while the ratio expected from the scaling relations is ~0.2-0.3%; McConnell & Ma 2013, Kormendy & Ho 2013). This is the case of e.g., the massive galaxies NGC 4291 and NGC 4342 (Bogdán et al. 2012), NGC 1332 (Rusli et al. 2011), SDSS J151741.75-004217.6 (Läsker et al. 2013), Mrk 1216

(Yıldırım et al. 2015; Ferré-Mateu et al. 2017), NGC 1271 (Walsh et al. 2015; Ferré-Mateu et al. 2015), the ultracompact dwarf galaxy M60-UCD1 (Seth et al. 2014), or the dwarf elliptical Was 49b (Secrest et al. 2017). While these overmassive BHs challenge the synchronized SMBH-galaxy growth paradigm, they can be explained by tidal stripping (e.g., Volonteri et al. 2008; Barber et al. 2016; Volonteri et al. 2016) or by a two-phase formation mechanism (Naab et al. 2009; Oser et al. 2012; Hilz et al. 2013; Barber et al. 2016). In the first case, galaxies formed on the e.g., $M_{\rm BH} - M_{\rm bulge}$ scaling relation; however, their stellar bodies where stripped by tidal interactions, leaving behind a galactic core with a SMBH that has become an overmassive BH. This could explain the presence of overmassive BHs in (ultracompact) dwarf galaxies (e.g., Volonteri et al. 2008; Mieske et al. 2013; Seth et al. 2014; Barber et al. 2016). The lost of stellar mass could also occur in massive galaxies by tidal stripping (ram-pressure stripping; e.g., Ginat et al. 2016) in the cluster environment, which could explain the overmassive BH initially proposed in e.g., NGC 1277 (e.g., van den Bosch et al. 2012; Emsellem 2013; Fabian et al. 2013; Yıldırım et al. 2015; Scharwächter et al. 2016; Walsh et al. 2016; also overmassive with respect to the fundamental plane, e.g., Scharwächter et al. 2016). However, the presence of an overmassive BH in NGC 1277 has been heavily disputed (e.g., Graham et al. 2016) and such stripping is not expected to occur in BCGs residing at the gravitational center of galaxy clusters.

In the two-phase formation scenario, the core of the galaxy formed rapidly in a first phase at $z \geq 2$ (when the $M_{\rm BH} - M_{\rm bulge}$ relation had a higher normalization; e.g., Jahnke et al. 2009; Decarli et al. 2010; Caplar et al. 2015; though see Trakhtenbrot et al. 2015 for an overmassive BH also at high-z). Since then, dry mergers with satellite galaxies add material to the outer parts of the galaxy, which grows in size and stellar mass, keeping the center almost unvaried (second phase; e.g., De Lucia & Blaizot 2007; Lidman et al. 2012; Zhang et al. 2016). Nonetheless, the finding of high levels of star formation in BCGs beyond $z \sim 1$ suggests that star formation could also contribute to (or dominate) the assembly of stellar mass (Webb et al. 2015b; McDonald et al. 2016). This two-phase formation scenario could explain the overmassive BHs found in massive galaxies (e.g., van den Bosch et al. 2012; Yıldırım et al. 2015; Walsh et al. 2015, 2016), among which BCGs. The few BCGs with dynamical BH mass measurements tend to be overmassive with respect to the $M_{\rm BH} - \sigma$ relation but appear more in agreement, with a large scatter, with the $M_{\rm BH} - L_{\rm V}$ and $M_{\rm BH} - M_{\rm bulge}$ relations (e.g., Lauer et al. 2007; McConnell et al. 2011a; McConnell & Ma 2013). Such deviation could thus be explained if BCGs grow via dry mergers, which increase the galaxy mass, luminosity and radius more than stellar velocity dispersion (e.g., Lauer et al. 2007; Naab et al. 2009; Oser et al. 2012; Volonteri & Ciotti 2013; but see Savorgnan & Graham 2015).

A third possibility is that BCGs form from the high-redshift ($z > 6$) seed BHs invoked to explain to existence of quasars with BH masses of more than $10^9$ $M_\odot$ when the Universe was only 0.8 Gyr old (e.g., Mortlock et al. 2011; Wu et al. 2015; Jiang et al. 2016; see Reines & Comastri 2016; Mezcua 2017 for recent reviews) and then grow via mergers following the standard dark matter hierarchical merging of haloes (e.g., Volonteri et al. 2003; Yoo et al. 2007; Natarajan & Treister 2009). In this case, BCGs could significantly exceed $M_{\rm BH} \sim 10^{11}$ $M_\odot$ by $z \sim 0$. However, several authors argue that there is a limit of a few $10^{10}$ $M_\odot$ on the mass a BH can reach, caused by BH self-regulation processes (e.g., Natarajan & Treister 2009), accretion disk fragmentation due to gravitational instabilities (King 2016), or the transition from a standard geometrically thin disk to an ADAF with which strong jets are associated (Inayoshi & Haiman 2016; Ichikawa & Inayoshi 2017). This limit could be breached in maximal spinning BHs ($a = 1$; e.g, King 2016; Inayoshi & Haiman 2016). The finding that the BH masses implied from the fundamental plane for some BCGs exceeds the maximum BH mass inferred from analytical models thus suggests that SMBHs in BCGs might be spinning at their maximum rate.

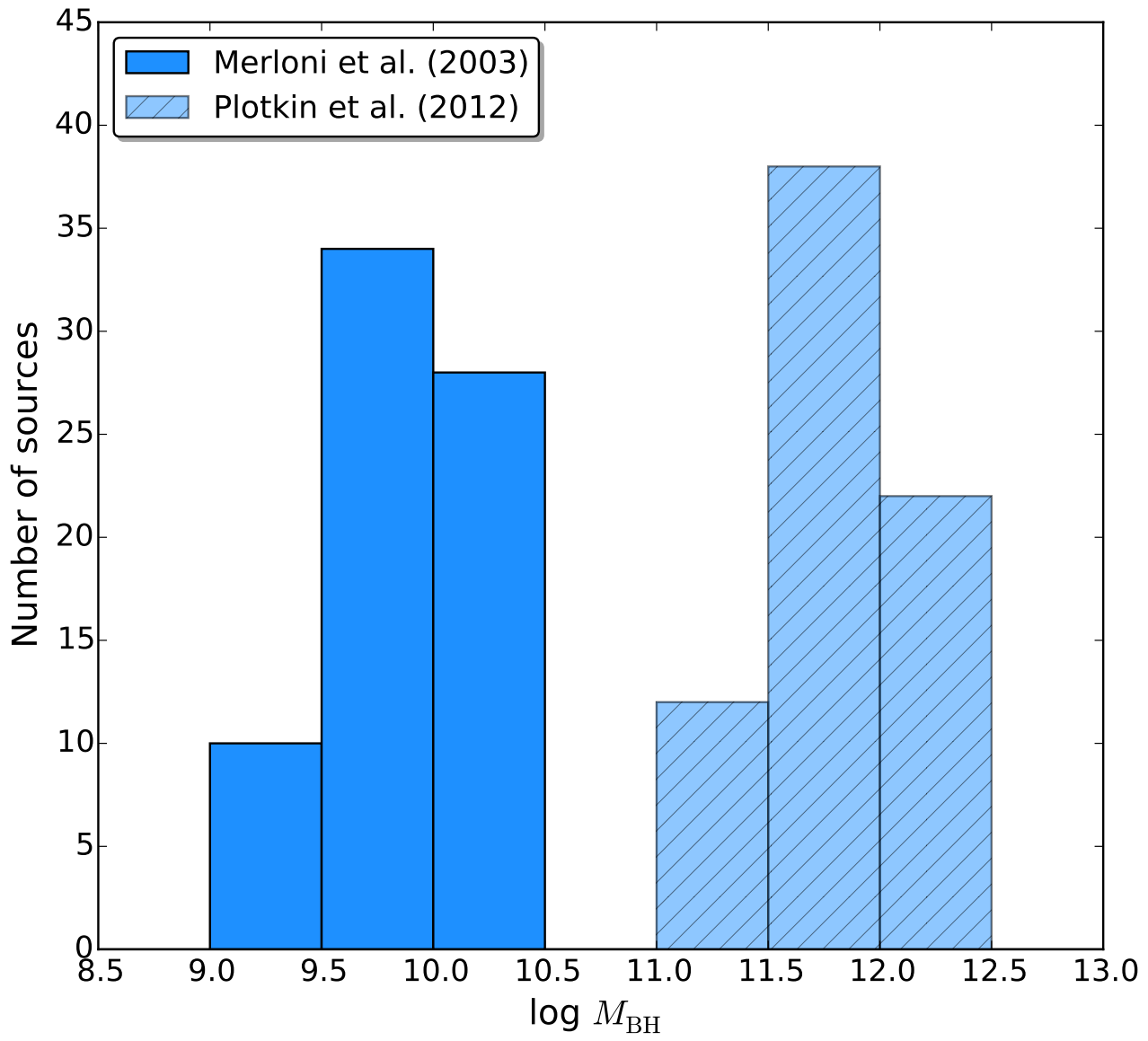

**Figure 12.** Distribution of BCG BH masses derived from the fundamental plane of Merloni et al. (2013, log $\Delta M_{\rm BH}$ = 0.3) and Plotkin et al. (2012, hashed bars, log $\Delta M_{\rm BH}$ = 2.2). Nearly 40% of the BCGs have $M_{\rm BH} > 10^{10}$ $M_\odot$ when considering the BH mass offset from Merloni et al. (2003), while all of them have $M_{\rm BH} > 10^{10}$ $M_\odot$ when considering that from Plotkin et al. (2012).

Both in the 'high-redshift seed' and 'two-phase' scenario, BCGs would have been accreting very rapidly and radiating efficiently as quasars at high redshifts, while in the local Universe they would become radiatively inefficient and have low accretion rates (i.e. because of gas depletion). Such a strong evolution was already suggested by Hlavacek-Larrondo et al. (2013) based on the finding that the fraction of radiatively efficient BCGs decreases from z~1 to z~0.1, and has been recently being supported by the high star formation rates found in some high-$z$ galaxy clusters (e.g., McDonald et al. 2012, 2013, 2016; Webb et al. 2015a,b).

### 4.7 Intracluster light

The cores of galaxy clusters are observed to be surrounded by a diffuse light made up of stars that are not gravitationally bound to any galaxy but to the cluster potential. This intracluster light (ICL) is thought to form from stars that have been tidally stripped during interactions and mergers between galaxies in the cluster (e.g., Gregg & West 1998; Conroy et al. 2007; Contini et al. 2014). In nearby clusters the ICL is found to be concentrated around the BCG, suggesting that the assembly and evolution of ICL and BCGs are closely linked. If this is the case, one may wonder whether the ICL could be contributing or should be added to the BCG stellar mass and therefore be affecting the BH masses estimated from the BH-galaxy scaling relations (such as the $M_{BH} - M_{\rm K}$ tested in this paper). Both observations and simulations show that the ICL has had its major growth at $z < 1$ (e.g., Monaco et al. 2006; Contini et al. 2014; Burke et al. 2015; Montes & Trujillo 2014; Morishita et al. 2017) and thus that it has been formed more recently than the BCG, whose stellar mass grew mainly until z=1 as after that most of the stellar mass from stripping and minor merging contributes to the ICL rather than the BCG (e.g., Conroy et al. 2007; Burke et al. 2015). Since our sample of BCGs spans only out to $z \sim 0.3$ and by then most of the BCG stellar mass is already in place, the ICL is not expected to affect the $M_{\rm K}$ measurement and thus nor to explain the presence of overmassive BHs in BCGs. We note though that the faint end of the BCG brightness profile often blends with the ICL, as finding an unambiguous definition of the transition between the inner bright region of the central galaxy and the low surface brightness where the ICL starts is very challenging (e.g., Pillepich et al. (2017)). Therefore, any small contribution of the ICL to the BCG stellar mass might be already included in the $M_{\rm K}$ measurement.

## 5 CONCLUSIONS

We have performed the first systematic study of the location of BCGs on the fundamental plane of BH accretion. For this we have used archival *Chandra* observations of a sample of 72 BCGs out to $z \sim 0.3$ and for which we have their 5 GHz core radio luminosity. Most of them (58 sources) are located in CC clusters. Nuclear X-ray emission is present in 41 of the BCGs while upper limits on the X-ray fluxes are derived for the remaining sources. All the BCGs have Eddington ratios $L_{\rm bol}/L_{\rm Edd} < 10^{-3}$ .

The nuclear X-ray luminosity is found to correlate with core radio luminosity as $L_{\rm X} \propto L_{\rm R}^{0.75\pm0.08}$, which favors an ADAF model over a jet for the origin of the X-ray emission. Assuming that BCGs follow the BH-galaxy scaling relations, we derive their BH masses from the $K$-band 2MASS magnitudes and the $M_{BH} - M_{\rm K}$ relation from Graham & Scott (2013). We then locate BCGs on the fundamental plane of BH accretion using this BH mass, the nuclear X-ray luminosity and the core radio luminosity. BCGs are found to be significantly offset from the fundamental plane of Plotkin et al. (2012) and Körding et al. (2006), which can be explained by the fact that these correlations favor a jet model, but also from the fundamental plane of Merloni et al. (2003), which favors an ADAF model as that found to originate the nuclear X-ray emission of BCGs. A significant offset is also obtained when considering other BH scaling relations. The offset is independent of whether BCGs are located in CCs or NCCs and thus on the amount of gas cooling. However, those BCGs located in CCs are found to have a higher duty cycle, which can be explained if cold accretion, in addition to Bondi accretion, takes place in CCs. Jet synchrotron cooling seems not to explain either the positive offset of BCGs from the fundamental plane. The offset from the fundamental plane is found to be higher for those core-dominated BCGs (whose radio emission comes predominantly from the core) in which the SMBH is currently actively accreting. These BCGs are found to have Eddington ratios on average ∼3 times higher than the weak-core BCGs (who have extended emission in addition to a weak core), as expected if the AGN in core-dominated BCGs accrete at higher rates because of being more currently active.

We derive the mass offset required for the BCGs to sit on average on the fundamental plane and find that it is, in the most conservative case (i.e. considering the fundamental plane of Merloni et al. 2003), of $\log \Delta \rm M_{BH} = 0.3$. Applying this mass 'correction' yields BH masses $\rm M_{BH} > 10^{10}$ $\rm M_{\odot}$ for nearly 40% of the BCGs in our sample, suggesting that they are ultramassive. The existence of ultramassive BHs in BCGs was already predicted by several analytical models (e.g., Natarajan & Treister 2009; King 2016) and confirmed for a few BCGs with dynamical BH mass measurements (e.g., McConnell et al. 2011a; McConnell & Ma 2013; see also Smith et al. 2017a,b). Hlavacek-Larrondo et al. (2012b) also suggested that strong CCs BCGs might host ultramassive BHs. By performing the most detailed study so far of the location of BCGs on the fundamental plane, we conclude that a large fraction of BCGs could host ultramassive BHs. These could either have formed via a two-phase formation process in which the core of the galaxy formed first (at high redshifts) and the rest of the galaxy grew later through dry mergers and/or in-situ star formation (e.g., Lauer et al. 2007; Volonteri & Ciotti 2013; Webb et al. 2015b; McDonald et al. 2016), or are the descendants of the high-redshift seed BHs required to explain the existence of quasars at $z > 6$ (e.g., Volonteri et al. 2003; Natarajan & Treister 2009). The finding that BCGs can host ultramassive BHs has thus important implications for the BH-galaxy scaling relations and the inferred SMBH-galaxy co-evolution paradigm.

## ACKNOWLEDGMENTS

The authors thank J.A Fernández-Ontiveros for the SED fitting. JHL is supported by NSERC through the discovery grant and Canada Research Chair programs, as well as FRQNT through the new university researchers start up program. JRL and ACE acknowledge support from STFC (ST/P000541/1). This publication makes use of data products from the Two Micron All Sky Survey, which is a joint project of the University of Massachusetts and the Infrared Processing and Analysis Center/California Institute of Technology, funded by the National Aeronautics and Space Administration and the National Science Foundation.

**Table 1.** Average BH mass offset from the fundamental plane and significance (in % and $\sigma$).

| Correlation | log $\Delta M_{BH}$ All | log $\Delta M_{BH}$ CCs | log $\Delta M_{BH}$ NCCs | log $\Delta M_{BH}$ core-dominated | log $\Delta M_{BH}$ weak-core | log $\Delta M_{BH}$ 150 GHz |
|---|---|---|---|---|---|---|
| Merloni et al. (2003) | 0.3 ± 1.2 | 0.3 ± 1.2 | 0.1 ± 1.1 | 0.5 ± 1.2 | 0.1 ± 1.2 | 1.3 ± 1.0 |
| | >98.0% (>2$\sigma$) | >98.3% (>2$\sigma$) | >59.0% (>0$\sigma$) | >99.1% (>2$\sigma$) | >65.1% (>0$\sigma$) | >99.9% (>3$\sigma$) |
| Plotkin et al. (2012) | 2.2 ± 1.6 | 2.3 ± 1.7 | 2.0 ± 1.6 | 2.4 ± 1.7 | 2.0 ± 1.6 | 3.4 ± 1.4 |
| | >99.9% (>8$\sigma$) | >99.9% (>8$\sigma$) | >99.9% (>4$\sigma$) | >99.9% (>6$\sigma$) | >99.9% (>6$\sigma$) | >99.9% (>5$\sigma$) |
| Plotkin et al. (2012) | 2.1 ± 2.1 | 2.1 ± 2.2 | 1.8 ± 2.0 | 2.3 ± 2.2 | 1.8 ± 2.2 | 3.3 ± 2.0 |
| (no cooling) | >99.9% (>7$\sigma$) | >99.9% (>6$\sigma$) | >99.7% (>3$\sigma$) | >99.9% (>5$\sigma$) | >99.9% (>4$\sigma$) | >99.9% (>4$\sigma$) |
| Gültekin et al. (2009b) | 0.1 ± 1.3 | 0.1 ± 1.3 | 0.0 ± 1.2 | 0.3 ± 1.2 | 0.0 ± 1.3 | 1.0 ± 1.1 |
| | >64.9% (>0$\sigma$) | >78.0% (>1$\sigma$) | >64.6% (>0$\sigma$) | >89.4% (>1$\sigma$) | >68.8% (>0$\sigma$) | >99.8% (>3$\sigma$) |
| Koerding et al. (2006) | 2.0 ± 1.5 | 2.1 ± 1.5 | 1.8 ± 1.4 | 2.2 ± 1.5 | 1.8 ± 1.5 | 3.3 ± 1.2 |
| | >99.9% (>8$\sigma$) | >99.9% (>8$\sigma$) | >99.9% (>4$\sigma$) | >99.9% (>7$\sigma$) | >99.9% (>6$\sigma$) | >99.9% (>5$\sigma$) |

**Table 2.** Sample X-ray properties

| Name | z | obsID | $N_{\rm H}$ ($10^{22}$ cm$^{-2}$) | $N_{\rm H,spec}$ ($10^{22}$ cm$^{-2}$) | Γ | log $L_{\rm phot,2-10keV}$ (erg s$^{-1}$) | log $L_{\rm spec,2-10keV}$ (erg s$^{-1}$) |
|---|---|---|---|---|---|---|---|
| A1204 | 0.1706 | 2205 | 0.0138 | 0 | 1.9 | 42.53±0.07 | $42.5^{+1}_{-0.2}$ |
| A1361 | 0.1167 | 2200 | 0.0224 | 0 | 1.9 | 41.56±0.2 | $41.4^{+0.5}_{-0.4}$ |
| A1367 | 0.0217 | 514 | 0.0239 | 0 | 1.9 | <39.36 | <39.7 |
| A1446 | 0.1028 | 4975 | 0.0150 | 0 | 1.9 | 41.67±0.04 | $41.4^{+0.3}_{-0.2}$ |
| A1644 | 0.0475 | 7922 | 0.0519 | 0 | 1.9 | <40.51 | <41.2 |
| A1664 | 0.1276 | 7901 | 0.0872 | 0 | 1.9 | <41.77 | <41.5 |
| A168 | 0.0443 | 3203 | 0.0325 | 0 | 1.9 | <40.64 | <40.2 |
| A1763 | 0.2280 | 3591 | 0.0092 | 0 | 1.9 | 42.21±0.09 | $42.3^{+0.3}_{-0.4}$ |
| A1795 | 0.0632 | 493,494 | 0.0117 | 0 | 1.9 | <41.27 | <40.8 |
| A1930 | 0.1316 | 11733 | 0.0113 | 0 | 1.9 | 41.37±0.2 | $41.9^{+0.2}_{-0.4}$ |
| A2009 | 0.1532 | 10438 | 0.0327 | 0 | 1.9 | 41.7±0.2 | $42.0^{+0.3}_{-0.4}$ |
| A2029 | 0.0780 | 4977 | 0.0315 | 0 | 1.9 | <41.86 | <41.2 |
| A2033 | 0.0780 | 15167 | 0.0294 | 0 | 1.9 | 41.26±0.1 | $41.6^{+0.5}_{-0.3}$ |
| A2052 | 0.0355 | 5807 | 0.0285 | 0 | $2.0^{+0.1}_{-0.1}$ | 41.26±0.01 | $41.0^{+0.09}_{-0.09}$ |
| A2063 | 0.0342 | 6262,6263 | 0.0304 | 0 | 1.9 | <40.21 | <40.1 |
| A2110 | 0.0976 | 15160 | 0.0239 | 0 | 1.9 | 41.3±0.1 | $41.8^{+0.3}_{-0.4}$ |
| A2199 | 0.0310 | 10748 | 0.0089 | 0 | 1.9 | <40.95 | <40.7 |
| A2204 | 0.1514 | 7940 | 0.0569 | 0 | 1.9 | <42.59 | <42.7 |
| A2355 | 0.2310 | 15097 | 0.0475 | 0 | 1.9 | 40.91±0.1 | $41.7^{+0.6}_{-0.3}$ |
| A2390 | 0.2328 | 4193 | 0.0689 | 0 | 1.9 | <42.22 | <42.3 |
| A2415 | 0.0573 | 12272 | 0.0479 | 0 | 1.9 | 41.69±0.06 | $41.3^{+0.3}_{-0.3}$ |
| A2597 | 0.0830 | 7329 | 0.0249 | 0 | 1.9 | <41.48 | <41.4 |
| A262 | 0.0166 | 7921 | 0.0546 | 0 | 1.9 | 39.82±0.04 | $39.4^{+0.2}_{-0.3}$ |
| A2626 | 0.0552 | 16136 | 0.0433 | 0 | 1.9 | 41.1±0.03 | $40.8^{+0.05}_{-0.06}$ |
| A2634 | 0.0298 | 4816 | 0.0506 | 0 | $1.4^{+0.3}_{-0.3}$ | 41.35±0.02 | $41.0^{+0.4}_{-0.2}$ |
| A2665 | 0.0567 | 12280 | 0.0604 | 0 | 1.9 | 40.86±0.1 | $41.3^{+0.2}_{-0.3}$ |
| A2667 | 0.2346 | 2214 | 0.0165 | 0 | 1.9 | 42.86±0.08 | $42.5^{+0.5}_{-0.4}$ |
| A3017 | 0.2195 | 15110 | 0.0209 | 0 | 1.9 | 42.4±0.09 | $42.5^{+0.5}_{-0.3}$ |
| A3526 | 0.0099 | 16223 | 0.0810 | 0 | 1.9 | <38.45 | <39.1 |
| A3528S | 0.0574 | 8268 | 0.0613 | 0 | 1.9 | <40.85 | <41.4 |
| A3581 | 0.0218 | 12884 | 0.0425 | $0.09^{+0.03}_{-0.02}$ | $2.2^{+0.1}_{-0.1}$ | 41.56±0.007 | $41.4^{+0.07}_{-0.06}$ |
| A3695 | 0.0888 | 12274 | 0.0370 | 0 | 1.9 | 41.82±0.07 | $42.3^{+0.2}_{-0.3}$ |
| A4059 | 0.0491 | 5785 | 0.0110 | 0 | 1.9 | <40.82 | <40.3 |
| A478 | 0.0860 | 1669 | 0.1508 | 0 | 1.9 | <41.47 | <41.3 |
| A496 | 0.0328 | 4976 | 0.0480 | 0 | 1.9 | <40.37 | <40.1 |
| AS1101 | 0.0564 | 11758 | 0.0183 | 0 | 1.9 | <40.76 | <40.7 |
| AS780 | 0.2344 | 9428 | 0.0772 | $6.0^{+4.0}_{-2.0}$ | 1.9 | 42.99±0.03 | $43.4^{+0.08}_{-0.08}$ |
| AS851 | 0.0095 | 11753 | 0.0496 | $2.0^{+0.9}_{-0.8}$ | 1.9 | 39.92±0.03 | $40.2^{+0.1}_{-0.2}$ |
| Hercules-A | 0.1550 | 5796,6257 | 0.0633 | 0 | 1.9 | <41.93 | <41.7 |
| Hydra | 0.0549 | 4970 | 0.0484 | $2.0^{+1.0}_{-0.7}$ | $1.4^{+0.8}_{-0.5}$ | 41.55±0.03 | $41.9^{+1}_{-0.2}$ |
| RXJ0058.9+2657 | 0.0480 | 6830 | 0.0573 | 0 | 1.9 | 41.06±0.03 | $39.4^{+0.4}_{-0.3}$ |
| RXJ0107.4+3227 | 0.0175 | 2147 | 0.0541 | 0 | $1.6^{+0.2}_{-0.3}$ | 40.94±0.02 | $40.7^{+0.3}_{-0.2}$ |
| RXJ0123.6+3315 | 0.0169 | 2882 | 0.0523 | 0 | 1.9 | 40.11±0.07 | $39.8^{+0.3}_{-0.3}$ |
| RXJ0341.3+1524 | 0.0290 | 4182 | 0.1617 | 0 | 1.9 | 40.84±0.05 | $40.5^{+0.2}_{-0.3}$ |
| RXJ0352.9+1941 | 0.1090 | 10466 | 0.1388 | $2.0^{+2.0}_{-1.0}$ | 1.9 | 42.5±0.03 | $42.9^{+0.07}_{-0.1}$ |
| RXJ0439.0+0520 | 0.2452 | 9369,9761 | 0.1030 | 0 | 1.9 | 42.68±0.09 | $42.6^{+0.5}_{-0.4}$ |
| RXJ0751.3+5012 | 0.0236 | 15170 | 0.0509 | 0 | 1.9 | <39.56 | <40.1 |
| RXJ0819.6+6336 | 0.1186 | 2199 | 0.0416 | 0 | 1.9 | 41.53±0.1 | $41.6^{+0.1}_{-0.1}$ |
| RXJ1050.4-1250 | 0.0154 | 3243 | 0.0450 | 0 | 1.9 | 39.44±0.3 | $40.0^{+0.2}_{-0.3}$ |
| RXJ1304.3-3031 | 0.0104 | 4998 | 0.0601 | 0 | 1.9 | <39.86 | <39.7 |
| RXJ1315.4-1623 | 0.0093 | 9399 | 0.0494 | $0.2^{+0.2}_{-0.1}$ | 1.9 | 39.7±0.03 | $39.6^{+0.07}_{-0.1}$ |
| RXJ1320.1+3308 | 0.0377 | 6941 | 0.0105 | 0 | $2.1^{+0.4}_{-0.5}$ | 41.28±0.02 | $40.9^{+0.2}_{-0.2}$ |
| RXJ1501.1+0141 | 0.0066 | 12952 | 0.0425 | 0 | $2.0^{+0.1}_{-0.1}$ | 39.28±0.02 | $39.1^{+0.2}_{-0.2}$ |
| RXJ1504.1-0248 | 0.2171 | 5793 | 0.0610 | 0 | 1.9 | <42.83 | <42.9 |
| RXJ1506.4+0136 | 0.0057 | 7923 | 0.0424 | 0 | 1.9 | 38.94±0.04 | $38.5^{+0.5}_{-0.2}$ |
| RXJ1522.0+0741 | 0.0451 | 900 | 0.0305 | 0 | 1.9 | <40.68 | <40.8 |

| Name | z | obsID | $N_H$ ($10^{20}$ cm$^{-2}$) | $N_{H,spec}$ ($10^{20}$ cm$^{-2}$) | $\Gamma$ | log $L_{phot,2-10keV}$ (erg s$^{-1}$) | log $L_{spec,2-10keV}$ (erg s$^{-1}$) |
|---|---|---|---|---|---|---|---|
| RXJ1524.2-3154 | 0.1022 | 9401 | 0.0844 | 0 | 1.9 | <42.02 | <41.7 |
| RXJ1539.5-8335 | 0.0758 | 8266 | 0.0768 | 0 | 1.9 | 41.36±0.2 | $41.7^{+0.5}_{-0.3}$ |
| RXJ1558.4-1410 | 0.0970 | 9402 | 0.1147 | 0 | 1.9 | 41.94±0.04 | $41.6^{+0.1}_{-0.2}$ |
| RXJ1604.9+2356 | 0.0324 | 9423 | 0.0499 | 0 | 1.9 | 40.45±0.05 | $39.9^{+0.3}_{-0.3}$ |
| RXJ1715.3+5725 | 0.0282 | 4194 | 0.0260 | 0 | $1.9^{+0.6}_{-0.7}$ | 40.71±0.06 | $40.5^{+1}_{-0.3}$ |
| RXJ1720.1+2638 | 0.1611 | 4361 | 0.0389 | 0 | 1.9 | <42.24 | <42.8 |
| RXJ1750.2+3504 | 0.1712 | 12252 | 0.0312 | 0 | 1.9 | 41.37±0.6 | $42.0^{+0.2}_{-0.4}$ |
| RXJ1844.1+4533 | 0.0917 | 5295 | 0.0632 | 0 | 1.9 | 42.16±0.03 | $41.9^{+0.2}_{-0.2}$ |
| RXJ2129.6+0005 | 0.2346 | 9370 | 0.0416 | 0 | 1.9 | <42.59 | <42.8 |
| Z1665 | 0.0311 | 15161 | 0.0274 | 0 | 1.9 | <40.86 | <40.9 |
| Z235 | 0.0830 | 11735 | 0.0391 | 0 | 1.9 | <41.58 | <41.3 |
| Z3146 | 0.2906 | 9371 | 0.0293 | 0 | 1.9 | <43.11 | <43.1 |
| Z7160 | 0.2578 | 4192 | 0.0322 | 0 | 1.9 | <42.45 | <43.3 |
| Z808 | 0.1690 | 12253 | 0.0755 | 0 | 1.9 | <42.42 | <42.3 |
| Z8193 | 0.1754 | 14988 | 0.0231 | 0 | 1.9 | <42.26 | <42.4 |
| Z8276 | 0.0750 | 11708 | 0.0366 | $0.7^{+0.6}_{-0.2}$ | $2.5^{+1.0}_{-0.6}$ | 41.98±0.02 | $41.9^{+0.9}_{-0.2}$ |

**Column designation:** (1) BCG name, (2) redshift, (3) *Chandra* obsID, (4) Galactic column density measured by Kalberla et al. (2005), (5) column density derived from the spectral fitting, (6) photon index, (7) 2-10 keV X-ray luminosity derived using the photometric method, (8) 2-10 keV X-ray luminosity derived from the spectral fitting.

**Table 3.** Sample luminosities and BH masses

| Name | log $L_{5\mathrm{GHz}}$ (erg s$^{-1}$) | $M_\mathrm{K}$ | log $M_{\mathrm{BH,K}}$ ($\mathrm{M}_\odot$) |
|---|---|---|---|
| A1204 | 39.86±0.07 | -26.3±0.1 | 9.49±0.02 |
| A1361 | 39.6±0.06 | -26.77±0.05 | 9.68±0.03 |
| A1367 | 39.88±0.07 | -25.38±0.01 | 9.07±0.01 |
| A1446 | 39.97±0.009 | -26.47±0.05 | 9.55±0.02 |
| A1644 | 40.46±0.003 | -27.21±0.03 | 9.88±0.04 |
| A1664 | 40.48±0.05 | -26.28±0.05 | 9.47±0.02 |
| A168 | 38.58±0.05 | -26.28±0.03 | 9.47±0.02 |
| A1763 | 40.46±0.06 | -27.8±0.1 | 10.12±0.05 |
| A1795 | 40.26±0.03 | -27.00±0.04 | 9.78±0.03 |
| A1930 | 40.11±0.04 | -27.5±0.2 | 10.02±0.05 |
| A2009 | 40.13±0.003 | -27.3±0.1 | 9.90±0.04 |
| A2029 | 39.37±0.1 | -27.70±0.02 | 10.09±0.05 |
| A2033 | 39.84±0.005 | -27.32±0.05 | 9.93±0.04 |
| A2052 | 40.69±0.09 | -26.71±0.02 | 9.66±0.03 |
| A2063 | 38.43±0.2 | -26.29±0.04 | 9.47±0.02 |
| A2110 | 39.34±0.02 | -26.66±0.05 | 9.64±0.03 |
| A2199 | 39.33±0.01 | -26.54±0.01 | 9.58±0.02 |
| A2204 | 40.57±0.02 | -27.27±0.04 | 9.90±0.04 |
| A2355 | 40.38±0.1 | -27.4±0.2 | 9.94±0.05 |
| A2390 | 42.11±0.02 | -27.7±0.3 | 10.11±0.06 |
| A2415 | 40.79±0.01 | -25.49±0.05 | 9.12±0.02 |
| A2597 | 40.48±0.1 | -25.80±0.05 | 9.26±0.02 |
| A262 | 38.0±0.01 | -25.96±0.01 | 9.33±0.02 |
| A2626 | 39.38±0.004 | -26.84±0.04 | 9.72±0.03 |
| A2634 | 40.42±0.009 | -26.32±0.01 | 9.48±0.02 |
| A2665 | 38.8±0.1 | -26.88±0.03 | 9.73±0.03 |
| A2667 | 40.9±0.2 | -26.8±0.2 | 9.70±0.04 |
| A3017 | 40.44±0.05 | -26.2±0.2 | 9.41±0.02 |
| A3526 | 39.02±0.04 | -26.33±0.01 | 9.49±0.02 |
| A3528S | 39.01±0.09 | -27.17±0.01 | 9.86±0.04 |
| A3581 | 40.29±0.04 | -25.6±0.02 | 9.17±0.01 |
| A3695 | 40.8±0.02 | -26.64±0.03 | 9.63±0.03 |
| A4059 | 38.83±0.06 | -27.38±0.03 | 9.95±0.04 |
| A478 | 39.83±0.02 | -26.71±0.03 | 9.66±0.03 |
| A496 | 39.79±0.02 | -26.91±0.02 | 9.75±0.03 |
| AS1101 | 39.1±0.07 | -26.68±0.05 | 9.65±0.03 |
| AS780 | 42.06±0.006 | -27.51±0.05 | 10.01±0.05 |
| AS851 | 39.06±0.009 | -25.91±0.01 | 9.31±0.02 |
| Hercules-A | 40.27±0.07 | -26.98±0.07 | 9.77±0.03 |
| Hydra | 41.04±0.008 | -26.33±0.03 | 9.49±0.02 |
| RXJ0058.9+2657 | 39.31±0.007 | -26.66±0.03 | 9.64±0.03 |
| RXJ0107.4+3227 | 39.46±0.002 | -26.01±0.01 | 9.35±0.02 |
| RXJ0123.6+3315 | 37.67±0.005 | -26.10±0.01 | 9.39±0.02 |
| RXJ0341.3+1524 | 39.06±0.02 | -24.86±0.02 | 8.84±0.02 |
| RXJ0352.9+1941 | 40.07±0.04 | -26.5±0.1 | 9.57±0.03 |
| RXJ0439.0+0520 | 42.21±0.07 | -27.13±0.03 | 9.84±0.04 |
| RXJ0751.3+5012 | 38.57±0.05 | -25.20±0.03 | 8.99±0.01 |
| RXJ0819.6+6336 | 38.9±0.09 | -27.19±0.04 | 9.87±0.04 |
| RXJ1050.4-1250 | 37.53±0.03 | -25.37±0.01 | 9.07±0.01 |
| RXJ1304.3-3031 | 38.42±0.1 | -25.83±0.01 | 9.27±0.02 |
| RXJ1315.4-1623 | 38.31±0.06 | -25.49±0.01 | 9.12±0.01 |
| RXJ1320.1+3308 | 39.14±0.09 | -25.54±0.03 | 9.14±0.01 |
| RXJ1501.1+0141 | 37.08±0.06 | -24.91±0.01 | 8.87±0.02 |
| RXJ1504.1-0248 | 41.41±0.008 | -26.7±0.06 | 9.65±0.03 |
| RXJ1506.4+0136 | 37.43±0.03 | -25.17±0.01 | 8.98±0.01 |
| RXJ1522.0+0741 | 38.76±0.01 | -25.96±0.04 | 9.33±0.02 |

| Name | log $L_{5GHz}$ (erg s$^{-1}$) | $M_K$ | log $M_{BH,K}$ ($M_\odot$) |
|---|---|---|---|
| RXJ1524.2-3154 | 40.58±0.005 | -26.57±0.06 | 9.6±0.03 |
| RXJ1539.5-8335 | 40.3±0.0006 | -27.14±0.07 | 9.85±0.04 |
| RXJ1558.4-1410 | 41.83±0.1 | -27.35±0.05 | 9.94±0.04 |
| RXJ1604.9+2356 | 39.67±0.004 | -26.51±0.02 | 9.57±0.02 |
| RXJ1715.3+5725 | 39.4±0.07 | -26.25±0.01 | 9.46±0.02 |
| RXJ1720.1+2638 | 40.15±0.02 | -27.2±0.1 | 9.86±0.04 |
| RXJ1750.2+3504 | 41.17±0.0003 | -27.3±0.05 | 9.91±0.04 |
| RXJ1844.1+4533 | 40.83±0.01 | -26.81±0.04 | 9.70±0.03 |
| RXJ2129.6+0005 | 40.62±0.05 | -27.28±0.03 | 9.91±0.04 |
| Z1665 | 38.96±0.05 | -26.01±0.03 | 9.35±0.02 |
| Z235 | 40.42±0.02 | -26.77±0.08 | 9.68±0.03 |
| Z3146 | 40.05±0.03 | -27.37±0.34 | 9.95±0.06 |
| Z7160 | 40.35±0.04 | -27.9±0.1 | 10.17±0.06 |
| Z808 | 39.78±0.1 | -26.77±0.19 | 9.68±0.03 |
| Z8193 | 41.61±0.005 | -27.58±0.02 | 10.04±0.05 |
| Z8276 | 40.68±0.02 | -26.60±0.07 | 9.61±0.03 |

**Column designation:** (1) BCG name, (2) 2-10 keV X-ray luminosity derived using the photometric method, (3) 5 GHz radio luminosity, (4) 2MASS $K$-band absolute magnitude of the bulge, (5) BH mass estimated from the $K$-band bulge luminosity using the correlation from Graham & Scott (2013).